\documentclass[aps,prd,onecolumn,12pt,nofootinbib,superscriptaddress,longbibliography,floatfix]{revtex4-2}

\usepackage{amsmath}
\usepackage{amssymb}
\usepackage{dcolumn}
\usepackage{bm}
\usepackage{graphicx}
\usepackage{float}
\usepackage{slashed}
\usepackage{setspace}
\usepackage{xcolor}
\usepackage[colorlinks=true,linkcolor=blue,citecolor=blue,urlcolor=blue]{hyperref}

\newcommand{\Jpsi}{J/\psi}
\newcommand{\psip}{\psi(2S)}
\newcommand{\etac}{\eta_{c}}
\newcommand{\etap}{\eta^{\prime}}
\newcommand{\etapp}{\eta^{(\prime)}}
\newcommand{\vsq}{\langle v^{2}\rangle}
\newcommand{\Rzero}{R_{\psi_{n}}(0)}
\newcommand{\eps}{\varepsilon}

\begin{document}

\title{Radiative decays of $J/\psi,\,\psi(2S)\rightarrow\gamma\etapp$
in perturbative QCD with relativistic corrections}

\author{Jun-Kang He}
\affiliation{College of Physics and Electronic Science, Hubei Normal University,
Huangshi 435002, China}

\author{Chao-Jie Fan}
\email{fancj@hbnu.edu.cn}
\affiliation{College of Physics and Electronic Science, Hubei Normal University,
Huangshi 435002, China}

\author{Cong Wang}
\email{wangj@hbnu.edu.cn}
\affiliation{College of Physics and Electronic Science, Hubei Normal University,
Huangshi 435002, China}

\begin{abstract}
We present the first calculation of the radiative decays $J/\psi,\psi(2S)\rightarrow\gamma\etapp$ in
perturbative QCD that includes the order-$q^{2}$ relativistic corrections in all three short-distance
contributions, namely the quark-antiquark, two-gluon, and QED contributions. The amplitudes are found to be remarkably insensitive to the light-cone distribution
amplitude and to the light-quark mass, a robustness that persists through order $q^{2}$ and makes the
predictions correspondingly reliable. The relativistic correction enhances the $J/\psi$ branching
ratios by roughly a factor of 2, narrowing their shortfall from experiment, whereas for the
$\psi(2S)$ it is about twice as large as for the $J/\psi$ and the low-order expansion converges
poorly. In two representative $\eta$--$\etap$ mixing schemes, the ratio
$\mathcal{R}_{1S}=\mathcal{B}(\gamma\etap)/\mathcal{B}(\gamma\eta)$ proves sharply sensitive to
the mixing angle and favors the smaller of the two.
The predicted $\psi(2S)$ rates lie well above the data in both channels, already at leading order,
and most severely for the anomalously small $\gamma\eta$ channel. Such a discrepancy suggests that
a mechanism beyond the hard perturbative process is at work. As a physically motivated attempt, we
explore the $\etac$-mixing contribution, which adds coherently to the perturbative one and is
comparable to it in the $\gamma\eta$ channel, and find that the interference can bring the
$\psi(2S)$ rates into agreement with the data, although its extraction is limited by a strong
sensitivity to the mixing parameters.
\end{abstract}

\maketitle

\section{Introduction}
\label{sec:intro}

The discovery of the $\Jpsi$~\cite{Aubert:1974js,Augustin:1974xw} opened the field of charmonium
physics and provided a uniquely clean window on the strong interaction. As the lowest $c\bar c$
vector state, the $\Jpsi$, together with its radial excitation $\psip$, lies in the transition
region between the perturbative and nonperturbative regimes of QCD, where the charm-quark mass sets
a semihard scale at which the strong coupling is neither small nor large and perturbative and
nonperturbative effects are closely intertwined~\cite{Brambilla:2004wf,Brambilla:2010cs,
Voloshin:2007dx,Eichten:2007qx}. The Okubo-Zweig-Iizuka- (OZI) forbidden radiative decays $\Jpsi,\psip\rightarrow\gamma\etapp$ are a prime manifestation
of this interplay. In the parton picture they proceed through the annihilation of the $c\bar c$
pair into a photon and at least two gluons, which subsequently convert into the
$\etapp$~\cite{Voloshin:2007dx,Eichten:2007qx}, so that they probe at once the conversion
of gluons into light hadrons, the gluonic content of the $\etapp$, and the
$\eta$--$\etap$ mixing that organizes the pseudoscalar nonet, the ratio of the two rates
being one of the classic observables for extracting the mixing angle.

Reflecting this perturbative-nonperturbative interplay, these decays have been described by two
distinct theoretical approaches. The first is nonperturbative, and several variants have been developed.
In the formulation of Novikov \textit{et al.}~\cite{Novikov:1979uy} the transition is assumed to be
dominated by the
$U_{A}(1)$ anomaly and is controlled by the gluonic matrix elements
$\langle0|G^{a}_{\mu\nu}\tilde G^{a,\mu\nu}|\etapp\rangle$, subsequently estimated in the
large-$N_{c}$ expansion and the QCD multipole formalism~\cite{Kuang:1990kd}. The
Feldmann-Kroll-Stech (FKS) scheme~\cite{Feldmann:1998vh} reexpresses these matrix elements
through the phenomenological constants $f_{q},f_{s}$ and the mixing angle $\phi$. In this picture
the anomaly form of the width carries a factor $\big(M_{J/\psi}/m_{c}^{2}\big)^{4}$
~\cite{Novikov:1979uy}, whose steep dependence on the charm mass renders the absolute normalization
strongly uncertain, while the mixing angle extracted from the radiative ratio under the assumption
of anomaly dominance, $\phi\simeq39^{\circ}$, is tied to that assumption. The same decays have also
been analyzed with effective-Lagrangian and vector-meson-dominance methods that describe the
$\Jpsi$ and $\psip$ transitions on a common footing~\cite{Gerard:2013gya,Zhao:2010mm}. A third,
closely related description regards the decay as proceeding through a small $\etac$ component
admixed into the $\etapp$. This $\etapp$--$\etac$ mixing picture was introduced by
Chao~\cite{Chao:1989pi,Chao:1990im} and adopted in the FKS analysis~\cite{Feldmann:1998vh} and is
in fact the same physics as the anomaly mechanism, since the gluonic $U_{A}(1)$ anomaly that couples
the $\etapp$ to two gluons is what mixes a $c\bar c$ component into it. We note, however,
that these $\etac$-mixing estimates commonly omit the radial overlap form factor associated with
probing the $\etac$ far off its mass shell. When it is neglected, the large phase-space factor
associated with the highly energetic photon can lead to a substantial overestimate of the
$\etac$-mixing contribution to these OZI-forbidden radiative decays, an enhancement that is likely
unphysical.

The second approach is perturbative. The amplitude for these decays is generated at leading twist by the
subprocess $c\bar c\rightarrow\gamma g^{*}g^{*}$, with the virtual gluons hadronizing into the
$\etapp$ through its light-cone distribution amplitudes (DAs). This perturbative QCD
approach was pioneered by K\"orner \textit{et al.}~\cite{Korner:1982vg} and by
K\"uhn~\cite{Kuhn:1983yr} and revisited by several groups using light-cone
DAs~\cite{Ma:2002ww,Yang:2004wy,Li:2005ug,Li:2007dq,Gao:2006}. In a recent
study~\cite{He:2019} we computed $\Jpsi\rightarrow\gamma\etapp$ in this framework keeping the light-
quark masses in the loop integrals and found that the sum of the one-loop integrals is infrared
finite and remarkably insensitive both to the light-quark masses and to the shape of the
$\etapp$ DAs. As a consequence the hard mechanism alone reproduces the measured ratio
$\mathcal{R}_{1S}=\mathcal{B}(J/\psi\rightarrow\gamma\etap)/\mathcal{B}(J/\psi\rightarrow\gamma\eta)$,
while the individual branching ratios come out somewhat below the data. The mixing angle extracted from $\mathcal{R}_{1S}$ in the hard picture,
$\phi\simeq33.5^{\circ}$, is appreciably smaller than the FKS value obtained under the assumption
of anomaly dominance. That analysis established that the hard mechanism is capable of describing
$\Jpsi\rightarrow\gamma\etapp$ at the level of present data.

Common to all of these treatments is that the initial charmonium is described in the
nonrelativistic, weak-binding approximation, in which the relative motion of the $c$ and $\bar c$
is neglected. Charmonium is, however, only moderately nonrelativistic, with a mean squared
heavy-quark velocity $\vsq\simeq0.3$, so corrections of relative order $v^{2}$ need not be small,
and their importance is well established across charmonium production and decay. A prominent
example on the production side is the exclusive process $e^{+}e^{-}\rightarrow\Jpsi+\etac$, whose
cross section measured at the $B$ factories exceeds the leading-order nonrelativistic QCD (NRQCD) prediction by nearly an
order of magnitude, a discrepancy that the relativistic corrections of order $v^{2}$, together with
the QCD radiative corrections, are essential to resolve~\cite{He:2007dc,Bodwin:2006dm}. On the
decay side, relativistic corrections have likewise been found to be substantial in exclusive
charmonium decays, both in the three-gluon decay $\Jpsi\rightarrow ggg$~\cite{He:2025Vggg} and in
exclusive hadronic channels such as $\Jpsi\rightarrow p\bar p$ and
$\psi(nS)\rightarrow\rho\pi$~\cite{Kivel:2022pp,Kivel:2023rp}. The underlying reason is that the
expansion parameter $v^{2}$ is not small, so the corrections of order $v^{2}$ can reach order
unity, in particular whenever the leading-order amplitude is itself suppressed. The radially
excited $\psip$ is even more sensitive, being a more loosely bound excited state whose mean squared
velocity is substantially larger than that of the $1S$ states~\cite{Kivel:2023rp,He:2026ggg2S}. The
$v^{2}$ expansion parameter is correspondingly larger, so the relativistic corrections are enhanced
and the expansion converges more slowly, an effect already seen in $\psi(2S)$
production~\cite{Elekina:2009qns}. Our recent study of the three-gluon decays
$\psi(2S),\Upsilon(2S)\rightarrow ggg$ reaches a similar conclusion~\cite{He:2026ggg2S}. In addition, the related $P$-wave radiative decays $h_{c}\rightarrow\gamma\etapp$ were also found
to receive significant relativistic corrections~\cite{He:2020}.

A reliable description of $\Jpsi,\psip\rightarrow\gamma\etapp$, and especially of the $\psip$
modes, therefore requires the relativistic corrections to the initial bound state to be included
systematically. We do so within the Bethe-Salpeter formalism. The present work provides, to the best of our knowledge, the
first calculation of these decays in which the $q^{2}$ relativistic corrections to the charmonium
are included consistently in all three short-distance contributions, namely the quark-antiquark,
two-gluon, and QED contributions, using the same covariant Salpeter projector and $q^{2}$ expansion as in our
three-gluon analyses~\cite{He:2025Vggg,He:2026ggg2S}, now convolved with the $\etapp$
light-cone DAs.

A further motivation comes from the $2S$ states. Experimentally the
$\psi(2S)\rightarrow\gamma\etapp$ rates are strongly suppressed:
$\mathcal{B}(\psi(2S)\rightarrow\gamma\etap)=(1.24\pm0.04)\times10^{-4}$ and
$\mathcal{B}(\psi(2S)\rightarrow\gamma\eta)=(9.2\pm1.8)\times10^{-7}$~\cite{BESIII:2017geta,
ParticleDataGroup:2024}, to be compared with
$\mathcal{B}(\Jpsi\rightarrow\gamma\etap)=(5.28\pm0.06)\times10^{-3}$ and
$\mathcal{B}(\Jpsi\rightarrow\gamma\eta)=(1.090\pm0.013)\times10^{-3}$~\cite{ParticleDataGroup:2024}. The naive
``$12\%$ rule'' follows from the measured leptonic widths and predicts that exclusive $\psi(2S)$ rates
be about $12\%$ of the corresponding $\Jpsi$ ones. Relative to this expectation, the $\gamma\etap$ mode
is suppressed by roughly a further factor of 5 and the
$\gamma\eta$ mode by more than 2 orders of magnitude. This pattern is reminiscent of the
long-standing $\rho\pi$ puzzle in $\psi(2S)$ decays~\cite{Brambilla:2010cs,Kivel:2023rp} and may share its
dynamical origin. Because the $2S$ radial wave function carries a node, a relativistic treatment
able to resolve the short-distance overlap is the natural tool to probe the $2S$ rates,
which makes $\Jpsi,\psip\rightarrow\gamma\etapp$ a particularly instructive testing ground for the
relativistic-correction framework.

The remainder of this paper is organized as follows. In Sec.~\ref{sec:framework} we set up the
calculation and derive the quark, gluon, and QED contributions in turn, including the relativistic
corrections. The numerical analysis is presented in Sec.~\ref{sec:numerical}, and we summarize in
Sec.~\ref{sec:summary}.

\section{Theoretical framework}
\label{sec:framework}

\subsection{Quark-content contribution}
\label{subsec:quark}

\begin{figure}[H]
\centering
\includegraphics[width=0.5\textwidth]{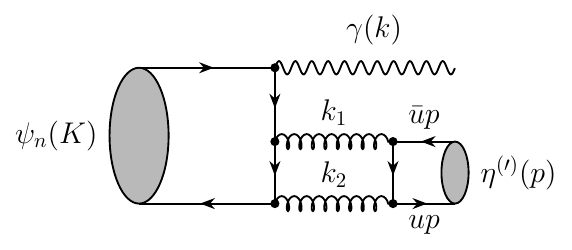}
\caption{A representative Feynman diagram for the quark-content contribution to
$\psi_{n}\rightarrow\gamma\etapp$, in which the $c\bar c$ pair annihilates into a photon and two
virtual gluons that couple to the $q\bar q$ content of the $\etapp$. Five further diagrams are
obtained by permuting the photon and gluon attachments on the charm line.}
\label{fig:quark}
\end{figure}

For the quark content of the $\etapp$, the decay proceeds in the parton picture through the
annihilation of the $c\bar c$ pair into a photon and two virtual gluons, which subsequently convert
into the $q\bar q$ content of the $\etapp$. At leading order this is a one-loop process, a
representative diagram of which is shown in Fig.~\ref{fig:quark}.

We work in the rest frame of the decaying charmonium $\psi_{n}$, where $n=1,2$ labels $\Jpsi$ and
$\psip$, and denote by $K$ the four-momentum of the charmonium of mass $M$, by $k$ that of the
real photon, and by $p=K-k$ that of the outgoing pseudoscalar of mass $m\equiv m_{\etapp}$. Following the treatment of K\"orner \textit{et al.}~\cite{Korner:1982vg},
which we also adopted in our previous nonrelativistic study of
$\Jpsi\rightarrow\gamma\etapp$~\cite{He:2019}, we evaluate the amplitude in two steps: the
$\psi_{n}\rightarrow\gamma g^{*}g^{*}$ and $g^{*}g^{*}\rightarrow\etapp$ amplitudes are computed separately
and then joined through the two gluon propagators and the loop integration over the gluon momentum.

We first consider the $\psi_{n}\rightarrow\gamma g^{*}g^{*}$ amplitude. In the Bethe-Salpeter framework it
is the four-dimensional convolution of the Bethe-Salpeter wave function $\Psi(K,q)$ with the hard kernel, the
$c$ and $\bar c$ carrying momenta $K/2\pm q$,
\begin{equation}
A=\sqrt{3}\int\!\frac{d^{4}q}{(2\pi)^{4}}\,\mathrm{Tr}\!\left[\Psi(K,q)\,\mathcal{O}(q)\right],
\label{eq:BSamp4d}
\end{equation}
where $\sqrt{3}$ is the color factor of the $c\bar c$ pair. The relative momentum of the pair is
decomposed into a component $q_{\parallel}$ longitudinal to $K$ and a transverse component
$\hat q$ with $\hat q\!\cdot\!K=0$,
\begin{equation}
q^{\mu}=q_{\parallel}^{\mu}+\hat q^{\mu},\qquad
q_{\parallel}^{\mu}=\frac{q\!\cdot\!K}{M^{2}}\,K^{\mu}.
\label{eq:qdecomp}
\end{equation}
In the rest frame of $\psi_{n}$, $K=(M,\bm{0})$ and the transverse component is purely spatial,
$\hat q=(0,\bm{\hat q})$, with $\bm{\hat q}$ the relative three-momentum.
For heavy quarkonium, whose constituents move nonrelativistically, the $c\bar c$ interaction can be
treated as instantaneous, so the Bethe-Salpeter kernel does not depend on the relative-energy
(longitudinal) component $q_{\parallel}$~\cite{Bhatnagar:2013bha,Gebrehana:2019}. In the same
nonrelativistic regime the relative energy is small compared with the heavy-quark mass, so the
constituent momenta reduce to $K/2\pm\hat q$ and the hard kernel $\mathcal{O}$ likewise depends only
on $\hat q$. The relative-energy dependence of $\Psi(K,q)$ is then integrated out over
$q_{\parallel}$, which defines the equal-time Salpeter wave function as the longitudinal projection
of the Bethe-Salpeter amplitude,
\begin{equation}
\psi(\hat q)=\frac{i}{2\pi}\int\!dq_{\parallel}\,\Psi(K,q),
\label{eq:salpeterdef}
\end{equation}
and reduces the amplitude to the three-dimensional
convolution~\cite{He:2026ggg2S,He:2025Vggg}
\begin{equation}
A=-i\sqrt{3}\int\!\frac{d^{3}\hat q}{(2\pi)^{3}}\,
\mathrm{Tr}\!\left[\psi(\hat q)\,\mathcal{O}(\hat q)\right].
\label{eq:BSamp}
\end{equation}
The Salpeter wave function of the $1^{--}$ state factorizes into a scalar radial function
$\chi(\hat q)$ and a Dirac projector $\mathcal{P}(\hat q)$ carrying the spin structure
\begin{equation}
\psi(\hat q)=\chi(\hat q)\,\mathcal{P}(\hat q).
\label{eq:salpeter}
\end{equation}
Being a scalar, $\chi(\hat q)$ can be pulled out of the Dirac trace and carries the entire
normalization. The Dirac structure of the $1^{--}$ wave function is organized by the power-counting
rule of Ref.~\cite{Bhatnagar:2013bha}, in which the covariants are ordered in powers of $1/M$.
Retaining the leading and next-to-leading covariants, that is up to first order in $\hat q$, all
Dirac structures reduce to a form sharing the single scalar function $\chi(\hat q)$, so that the
spin structure is carried entirely by the projector $\mathcal{P}(\hat q)$ while $\chi(\hat q)$
controls the convolution. The explicit reduction for the $1^{--}$ state is
given in Ref.~\cite{Gebrehana:2019}. The projector $\mathcal{P}(\hat q)$ then reads
\begin{equation}
\mathcal{P}(\hat q)=M\slashed{\eps}-\slashed{\eps}\slashed{K}
+\frac{M}{m_{c}}(\hat q\!\cdot\!\eps)-\frac{1}{m_{c}}\slashed{K}(\hat q\!\cdot\!\eps)
+\frac{1}{m_{c}}\slashed{K}\slashed{\eps}\,\slashed{\hat q},
\label{eq:proj}
\end{equation}
where $\eps\equiv\eps(K)$ is the charmonium polarization vector with $\eps\!\cdot\!K=0$. The first
two terms, $M\slashed{\eps}-\slashed{\eps}\slashed{K}$, give the leading ${}^{3}S_{1}$ component in
the nonrelativistic limit, while the terms linear in $\hat q$ are the subleading Dirac covariants
induced by the relative motion of the $c\bar c$ pair. The next covariants, of order $\hat q^{2}$,
are suppressed by a further power of $1/M$ in this counting~\cite{Bhatnagar:2013bha} and are
dropped. This truncation is sufficient for the $q^{2}$ relativistic corrections retained here. The
dynamical correction is absorbed entirely into the soft scalar function $\chi(\hat q)$, which is
kept in full. The kinematical correction originates from the internal-momentum dependence of the
hard-kernel denominators and becomes significant only when the internal momentum drives a
propagator toward its threshold~\cite{He:2025Vggg}. The $\hat q$ dependence carried by the Dirac
projector enters only the numerator, where it is immaterial to the kinematical correction. For the
$\psi(2S)$ the larger $\vsq$ enhances these corrections, and their convergence is examined in
Sec.~\ref{sec:numerical}.

The hard kernel $\mathcal{O}(\hat q)$ is the amplitude for $c\bar c\rightarrow\gamma g^{*}g^{*}$, in which
the photon and the two gluons are attached to the charm line in all six orderings. With the
charm momenta $K/2\pm\hat q$, the hard kernel reads
\begin{equation}
\mathcal{O}(\hat q)=iQ_{c}e\,g_{s}^{2}\,\frac{\delta^{ab}}{6}\,
\slashed{\eps}^{*}(k_{2})\,S_{F}(l_{2})\,\slashed{\eps}^{*}(k_{1})\,S_{F}(l_{1})\,
\slashed{\eps}^{*}(k)+(\text{5 permutations}),
\label{eq:hardkernel}
\end{equation}
where $\eps^{*}(k)$ is the photon polarization vector, $\eps^{*}(k_{1})$ and $\eps^{*}(k_{2})$ the
gluon polarization vectors, $a$ and $b$ the gluon color indices, and
$S_{F}(l)=(\slashed{l}+m_{c})/(l^{2}-m_{c}^{2})$ the two charm propagators, whose momenta along the
upper and lower charm lines are $l_{1}=K/2+\hat q-k$ and $l_{2}=K/2+\hat q-k-k_{1}$.

Since $\chi(\hat q)$ is sharply peaked at $\hat q\simeq0$, the trace
$\mathrm{Tr}[\mathcal{P}(\hat q)\mathcal{O}(\hat q)]$ is Taylor expanded to second order in
$\hat q$,
\begin{equation}
\mathrm{Tr}[\mathcal{P}(\hat q)\mathcal{O}(\hat q)]=\mathcal{T}_{0}+\hat q^{\mu}\mathcal{T}_{\mu}
+\hat q^{\mu}\hat q^{\nu}\mathcal{T}_{\mu\nu}+O(\hat q^{3}),
\label{eq:taylor}
\end{equation}
where
\begin{equation}
\mathcal{T}_{0}=\mathrm{Tr}[\mathcal{P}(0)\mathcal{O}(0)],\qquad
\mathcal{T}_{\mu}=\frac{\partial\,\mathrm{Tr}[\mathcal{P}(\hat q)\mathcal{O}(\hat q)]}
{\partial\hat q^{\mu}}\bigg|_{\hat q=0},\qquad
\mathcal{T}_{\mu\nu}=\frac{1}{2}\,\frac{\partial^{2}\,\mathrm{Tr}[\mathcal{P}(\hat q)\mathcal{O}(\hat q)]}
{\partial\hat q^{\mu}\,\partial\hat q^{\nu}}\bigg|_{\hat q=0}.
\label{eq:Tdef}
\end{equation}
The linear term vanishes by parity. Inside the convolution $\int d^{3}\hat q\,\chi(\hat q)\,(\cdots)$
the scalar function $\chi(\hat q)$ depends only on $|\bm{\hat q}|$ and is isotropic, so only the angular
average of $\hat q^{\mu}\hat q^{\nu}$ over the directions of $\hat q$ survives. This average is
\begin{equation}
\langle\hat q^{\mu}\hat q^{\nu}\rangle_{\rm dir}=|\bm{\hat q}|^{2}\,P_{T}^{\mu\nu},\qquad
P_{T}^{\mu\nu}=\frac{1}{3}\Big(-g^{\mu\nu}+\frac{K^{\mu}K^{\nu}}{M^{2}}\Big),
\label{eq:PT}
\end{equation}
where the transverse projector $P_{T}^{\mu\nu}$ projects onto the three-dimensional subspace
orthogonal to $K$, the factor $1/3$ being the isotropic average over the three spatial directions in
the rest frame. This replacement is valid only under the $\hat q$ integration and is a consequence
of the isotropy of $\chi(\hat q)$. The directionally averaged trace is then
\begin{equation}
\big\langle\mathrm{Tr}[\mathcal{P}(\hat q)\mathcal{O}(\hat q)]\big\rangle_{\rm dir}
=\mathcal{T}_{0}+|\bm{\hat q}|^{2}\,P_{T}^{\mu\nu}\mathcal{T}_{\mu\nu}+O(\hat q^{4}).
\label{eq:angavg}
\end{equation}
Inserting this into the three-dimensional convolution and carrying out the
$\hat q$ integration, the bound state enters only through its two moments at the origin,
\begin{equation}
\int\!\frac{d^{3}\hat q}{(2\pi)^{3}}\,\chi(\hat q)=\sqrt{\frac{1}{4M}}\,\sqrt{\frac{1}{4\pi}}\,\Rzero,
\qquad
\int\!\frac{d^{3}\hat q}{(2\pi)^{3}}\,|\bm{\hat q}|^{2}\,\chi(\hat q)
=-\sqrt{\frac{1}{4M}}\,\sqrt{\frac{1}{4\pi}}\,\nabla^{2}R_{\psi_{n}}(0),
\label{eq:moments}
\end{equation}
where $\sqrt{1/(4M)}$ arises from the relativistic normalization of the scalar function
$\chi(\hat q)$ and $\sqrt{1/(4\pi)}$ is the $S$-wave spherical harmonic. Retaining the relative-momentum expansion through the $q^{2}$
relativistic correction, the $\psi_{n}\rightarrow\gamma g^{*}g^{*}$ amplitude takes the compact form
\begin{equation}
A=-i\sqrt{3}\,\sqrt{\frac{1}{4M}}\,\sqrt{\frac{1}{4\pi}}\,\Big[\,R_{\psi_{n}}(0)\,\mathcal{T}_{0}
-\nabla^{2}R_{\psi_{n}}(0)\,P_{T}^{\mu\nu}\mathcal{T}_{\mu\nu}\,\Big].
\label{eq:Aintegrated}
\end{equation}
The first term, governed by the wave function at the origin $R_{\psi_{n}}(0)$, is the leading-order
amplitude, and the second, governed by its Laplacian $\nabla^{2}R_{\psi_{n}}(0)$, is the $q^{2}$
relativistic correction. The neglected $O(q^{4})$ terms are set by the higher moments of the
wave function.

The two virtual gluons are then coupled to the $\etapp$ through its quark content. The quark-antiquark content of
the $\etapp$ is described by the twist-2 light-cone matrix
element~\cite{Chernyak:1983ej}
\begin{equation}
\langle\etapp(p)|\bar q_{\alpha}(x)q_{\beta}(y)|0\rangle
=\frac{i}{4}f_{\etapp}^{q}\,(\slashed{p}\gamma_{5})_{\beta\alpha}
\int_{0}^{1}\!du\,e^{i(\bar u p\cdot y+u p\cdot x)}\phi^{q}(u)+\cdots,
\label{eq:quarkMT}
\end{equation}
where the higher-twist contributions are omitted and $f^{q}_{\etapp}$ is the flavor decay
constant of the $\etapp$. The flavor decay constants of the $\eta$ and $\etap$ are not independent.
In the quark-flavor basis of the $\eta$--$\etap$ system with a single mixing angle~\cite{Feldmann:1998vh}
they are parametrized by the two decay constants $f_{q}$ and $f_{s}$ and the mixing angle $\phi$,
whose values we specify in Sec.~\ref{subsec:inputs}. The twist-2 DA is
\begin{equation}
\phi^{q}(u)=6u\bar u\Big[1+\!\!\sum_{n=2,4,\dots}\!\!c^{q}_{n}(\mu)C^{3/2}_{n}(2u-1)\Big],
\label{eq:quarkDA}
\end{equation}
with $\bar u=1-u$ and $u$ the momentum fraction carried by the quark and $c^{q}_{n}(\mu)$ the
scale-dependent Gegenbauer coefficients. Because the $\etapp$ carries a flavor-singlet component,
the quark DA does not evolve autonomously. Under QCD evolution its coefficients $c^{q}_{n}(\mu)$ mix
with those of the two-gluon DA introduced in Sec.~\ref{subsec:gluon}, and the
coupled Efremov-Radyushkin-Brodsky-Lepage (ERBL) evolution equations and the associated anomalous dimensions are given in
Ref.~\cite{Kroll:2002nt}. With the quark content of the $\etapp$ thus specified, the two virtual
gluons couple to it by attaching to the light $q\bar q$ pair. Replacing that pair by the matrix
element of Eq.~(\ref{eq:quarkMT}) and convolving the hard $g^{*}g^{*}\rightarrow q\bar q$ subprocess with
the DA over the momentum fraction $u$ gives the effective
$g^{*}g^{*}\rightarrow\etapp$ vertex
\begin{equation}
M^{\mu\nu}=-i\,(4\pi\alpha_{s})\,\delta_{ab}\,\epsilon^{\mu\nu\rho\sigma}k_{1\rho}k_{2\sigma}
\sum_{q=u,d,s}\frac{f^{q}_{\etapp}}{6}\int_{0}^{1}\!du\,
\left(\frac{\phi^{q}(u)}{\bar u k_{1}^{2}+u k_{2}^{2}-u\bar u m^{2}-m_{q}^{2}}+(u\leftrightarrow\bar u)\right),
\label{eq:ggeta}
\end{equation}
with $k_{1},k_{2}$ the gluon momenta, $m_q$ the light-quark $q$ mass, and $m$ the $\etapp$ mass.

The two subamplitudes are now joined through the gluon lines. Since the gluons are internal, their
polarization vectors are replaced by the gluon propagators, and the upper
$\psi_{n}\rightarrow\gamma g^{*}g^{*}$ amplitude of Eq.~(\ref{eq:Aintegrated}), stripped of those
polarization vectors, enters as the tensor $A^{\mu\nu}$. Contracting it with the vertex $M_{\mu\nu}$
through the two propagators and integrating over the gluon loop momentum yields the decay amplitude
\begin{equation}
M_{T}=\frac{1}{2}\int\!\frac{d^{4}k_{1}}{(2\pi)^{4}}\,A^{\mu\nu}M_{\mu\nu}\,
\frac{i}{k_{1}^{2}}\,\frac{i}{k_{2}^{2}},
\label{eq:assemble}
\end{equation}
with $k_{2}=p-k_{1}$ and the factor $1/2$ compensating the interchange of the two identical gluons already summed in
$A^{\mu\nu}$ and $M^{\mu\nu}$. Let $T^{\alpha\beta}$ denote $M_{T}$ with the polarization vectors of
the $\psi_{n}$ and the photon removed. Lorentz covariance, parity, and gauge invariance fix it to the
single parity-odd structure $T^{\alpha\beta}\propto\,\epsilon^{\alpha\beta\mu\nu}K_{\mu}k_{\nu}$, so the
decay $\psi_{n}(1^{--})\rightarrow\gamma(1^{-})+\etapp(0^{-+})$ has one independent helicity
amplitude, extracted with the projector~\cite{Korner:1982vg}
\begin{equation}
h^{\alpha\beta}=-\frac{i}{2\,(K\!\cdot\!k)}\,\epsilon^{\alpha\beta\mu\nu}K_{\mu}k_{\nu},
\label{eq:helproj}
\end{equation}
normalized so that its contraction with $T^{\alpha\beta}$ returns the physical helicity amplitude,
\begin{equation}
H^{q}=h^{\alpha\beta}T_{\alpha\beta}.
\label{eq:helamp}
\end{equation}
Taking the Dirac trace, coupling the two gluons to the $\etapp$ through
Eq.~(\ref{eq:ggeta}), assembling the loop as in Eq.~(\ref{eq:assemble}), and summing the six
diagrams together with the flavor sum and the loop integration, the quark helicity amplitude takes
the explicit form
\begin{equation}
H^{q}=\frac{2Q_{c}}{2\sqrt{3}}\sqrt{4\pi\alpha_{e}}\,(4\pi\alpha_{s})^{2}\sqrt{\frac{1}{16\pi M}}\,
\sum_{q=u,d,s}f^{q}_{\etapp}\int_{0}^{1}\!du\,\phi^{q}(u)\Big[\,R_{\psi_{n}}(0)\,I^{q}_{0}(u)
-\nabla^{2}R_{\psi_{n}}(0)\,I^{q}_{2}(u)\,\Big],
\label{eq:Hq}
\end{equation}
with $Q_{c}=2/3$ the charm electric charge. The leading-order loop kernel is
\begin{equation}
I^{q}_{0}(u)=\int\!\frac{d^{4}k_{1}}{(2\pi)^{4}}\,
\frac{1}{G_{1}\,G_{2}\,L}
\left[\frac{N_{A}}{P_{0}\,P_{1}}+\frac{N_{B}}{P_{0}\,P_{2}}+\frac{N_{C}}{P_{1}\,P_{2}}\right]+(u\leftrightarrow\bar u).
\label{eq:integq0}
\end{equation}
The five propagators are, with $k_{1}$ the loop momentum and $k_{2}=p-k_{1}$,
$G_{1}=k_{1}^{2}+i\epsilon$ and $G_{2}=k_{2}^{2}+i\epsilon$ for the two gluon lines,
$L=(k_{1}-up)^{2}-m_{q}^{2}+i\epsilon$ for the light-quark line at momentum fraction $u$,
$P_{0}=(k-K/2)^{2}-m_{c}^{2}+i\epsilon$ for the loop-momentum-independent charm propagator on the
photon-emitting end of the charm line, and $P_{1}=(k_{1}-K/2)^{2}-m_{c}^{2}+i\epsilon$,
$P_{2}=(k_{1}+k-K/2)^{2}-m_{c}^{2}+i\epsilon$ for the two
loop-momentum-dependent charm propagators between consecutive gluon-charm vertices. The
six Feynman diagrams generated by permuting the photon and the two gluon attachments on the
charm line are pairwise equal under the interchange of the identical gluons, collapsing to
three independent topologies $X\in\{A,B,C\}$, namely two four-point box configurations ($X=A,B$,
with the photon attached at either end of the charm line), in which $P_{0}$ factors out of the
loop, and a five-point pentagon ($X=C$, with the photon attached between the two gluons), in
which both $k_{1}$-dependent charm lines remain off shell. The $(u\leftrightarrow\bar u)$
term in Eq.~(\ref{eq:integq0}) accounts for the second orientation of the light-quark loop,
obtained through $u\rightarrow\bar u=1-u$ in $L$ alone, since $L$ is the only propagator
that depends on $u$. Throughout the hard kernel we adopt the weak-binding approximation
$m_{c}=M/2$. It is consistent with the treatment of the Salpeter wave function above, whose spinor
structure is retained only to first order in $\hat q/M$. Our analyses of the three-gluon
decays~\cite{He:2025Vggg,He:2026ggg2S}, built on the same projector and $q^{2}$ expansion, adopt it
likewise. The three numerators are then polynomials in the scalar products
\begin{align}
N_{A}={}& 32 i\,M\,\Big[(k\!\cdot\!k_{1})^{2} - (k\!\cdot\!k_{1})\,(k_{1}\!\cdot\!K) + k_{1}^{2}\,k\!\cdot\!K\,\Big].
\label{eq:NA}
\end{align}
\begin{align}
N_{B}={}& 32 i\,M\,\Big[(k\!\cdot\!k_{1})^{2} - (k\!\cdot\!k_{1})\,(k_{1}\!\cdot\!K) - (k_{1}\!\cdot\!K)\,(k\!\cdot\!K) \nonumber \\
    &\quad  - (k\!\cdot\!k_{1})\,(k\!\cdot\!K) + k_{1}^{2}\,k\!\cdot\!K +M^2\,k\!\cdot\!k_{1}\,\Big].
\label{eq:NB}
\end{align}
\begin{align}
N_{C}={}& 32 i\,M\,\Big[2\,(k\!\cdot\!k_{1})\,(k_{1}\!\cdot\!K) - \frac{M^2\,(k\!\cdot\!k_{1})^{2}}{k\!\cdot\!K} - k_{1}^{2}\,k\!\cdot\!K\,\Big].
\label{eq:NC}
\end{align}
The forms quoted here are the leading term of the relative-momentum expansion, equal to the hard
kernel evaluated at $\hat q=0$, where the $c$ and $\bar c$ each carry half of the $\psi_{n}$
momentum. The split into topologies $A$,
$B$, and $C$ corresponds to the three positions of the photon vertex along the charm line relative to
the two gluon vertices. Individually they are not gauge invariant, the photon Ward identity being
realized only by the telescoping sum over insertion points whose boundary terms vanish for the
on-shell charm pair of the weak-binding projection. Only the sum over $X\in\{A,B,C\}$ is physical, and
it reproduces the compact loop function of Ref.~\cite{He:2019}. In addition, Bose symmetry under the
interchange of the two gluons, $k_{1}\rightarrow p-k_{1}$, relates the two boxes through
$N_{B}=N_{A}(k_{1}\rightarrow p-k_{1})$ and maps $N_{C}$ onto itself.

Beyond this leading order, the $q^{2}$-correction kernel $I^{q}_{2}(u)$ is generated by restoring the
relative-momentum dependence in the charm-line momenta as $K/2\pm\hat q$ and taking the
spherical average of the second derivative at $\hat q=0$ according to
Eq.~(\ref{eq:angavg}). Because $\hat q$ enters the loop integrand only through the
charm-line momenta, the second-derivative operation acts exclusively on the three
charm propagators $P_{0}(\hat q)$, $P_{1}(\hat q)$, and $P_{2}(\hat q)$. The two gluon
propagators $G_{1,2}$ and the light-quark propagator $L$ are inert under
$\partial_{\hat q}$ and remain in $I^{q}_{2}(u)$ in the very same form as in $I^{q}_{0}(u)$. Each
application of the Laplacian on a charm propagator either raises its multiplicity by one
or shifts a $P_{0}$ into a $P_{1}$ or $P_{2}$, so the charm-line denominator structure of $I^{q}_{2}(u)$
is considerably richer than at leading order. The result is organized by the powers $(a,b)$ of the charm denominators, with
$D^{A}_{a,b}=P_{0}^{a}P_{1}^{b}$ and $D^{B}_{a,b}=P_{0}^{a}P_{2}^{b}$ for the two boxes and
$D^{C}_{a,b}=P_{1}^{a}P_{2}^{b}$ for the pentagon, the powers running over $a,b\ge1$ with $a+b\le4$ to
give six distinct denominators per topology and 18 in all. Keeping these denominators in their derivative
form, rather than placing them over a common denominator, makes the Passarino-Veltman type of each
term manifest. With the same $(G_{1}G_{2}L)^{-1}$ skeleton inherited from leading order,
\begin{equation}
I^{q}_{2}(u)=\int\!\frac{d^{4}k_{1}}{(2\pi)^{4}}\,\frac{1}{G_{1}\,G_{2}\,L}
\sum_{X\in\{A,B,C\}}\,\sum_{(a,b)}\,\frac{N^{X}_{a,b}(k_{1})}{D^{X}_{a,b}}
+(u\leftrightarrow\bar u).
\label{eq:integq2}
\end{equation}
For the three lowest, undifferentiated $(a,b)=(1,1)$ terms the numerators take the
compact factorized forms
\begin{align}
N^{A}_{1,1}=& \,\frac{32i}{3 M}\Big[2\,k_{1}\!\cdot\!(k+K)\,(k\!\cdot\!K) + M^2\,k_{1}\!\cdot\!p - \frac{ M^4\,k\!\cdot\!k_{1}}{k\!\cdot\!K}\,\Big],
\label{eq:NXA11}
\end{align}

\begin{align}
N^{C}_{1,1}=& \frac{64 i}{3 M}\Big[ 4\,(k\!\cdot\!k_{1})\,(k_{1}\!\cdot\!K) + \frac{2 M^2\,(k\!\cdot\!k_{1})\,(k_{1}\!\cdot\!K)}{k\!\cdot\!K}-\,2\,(k_{1}\!\cdot\!K)^{2}  \nonumber \\
    &\quad  - \frac{3 M^2\,(k\!\cdot\!k_{1})^{2}}{k\!\cdot\!K} - \,k_{1}\!\cdot\!(k+K)\,(k\!\cdot\!K) - 3\,k_{1}^{2}\,k\!\cdot\!K\nonumber \\
    &\quad  +  M^2\,k_{1}\!\cdot\!K +2 M^2\,k\!\cdot\!k_{1}   -\frac{ M^4\,k\!\cdot\!k_{1}}{k\!\cdot\!K}\,\Big],
\label{eq:NXC11}
\end{align}
with $N^{B}_{1,1}=N^{A}_{1,1}(k_{1}\rightarrow p-k_{1})$. The remaining 15 numerators with $a+b>2$ are polynomials of total degree up to 3
in the same loop scalar products.
Their explicit forms, generated symbolically from the $\hat q^{2}$ derivative and verified
by reconstruction against the original integrand, are collected in
the Appendix. The subsequent numerical evaluation proceeds in two stages. The five
propagators of the pentagon topology are linearly dependent in the loop momentum $k_{1}$, so the
pentagon is first reduced to four- and lower-point integrals by partial fractioning over $k_{1}$
with the \texttt{ApartFF} routine of \textsc{FeynCalc}~\cite{Mertig:1990an,Shtabovenko:2016sxi,Shtabovenko:2020gxv}.
The resulting four-, three- and two-point tensor integrals are then passed to
\textsc{Package-X}~\cite{Patel:2015tea,Patel:2016fam}, which performs the tensor reduction of the
numerators and expresses the amplitude through the standard scalar Passarino-Veltman functions.

The leading-order analysis of Ref.~\cite{He:2019} already noted a striking empirical
feature of the loop kernel. $I^{q}_{0}(u)$ is, to very good accuracy, independent of both
the momentum fraction $u$ and the light-quark mass $m_{q}$ across their physically
relevant ranges. That observation was reported there but only briefly commented upon, and
its true origin was not identified. The present analysis brings out the same
near constancy in the $q^{2}$-correction kernel $I^{q}_{2}(u)$, a result by no means
obvious in advance given the much richer charm-line denominator structure of $I^{q}_{2}(u)$
relative to $I^{q}_{0}(u)$. The numerical study of Sec.~\ref{sec:numerical} confirms that
under independent variations of the DA $\phi^{q}(u)$ and of $m_{q}$
across the full physical interval the two kernels move by at most a few percent. The
recurrence of this property at the relativistic level is too systematic to be coincidental
and traces back to a sharp interplay between the analytic form of the integrands in
Eqs.~(\ref{eq:integq0}) and~(\ref{eq:integq2}) and the hard-mechanism character of the
underlying decay.

The analytic content of this interplay is most clearly seen at the integrand level. By
construction $u$ and $m_{q}$ appear in the integrands of Eqs.~(\ref{eq:integq0})
and~(\ref{eq:integq2}) solely
through the single light-quark propagator $L=(k_{1}-up)^{2}-m_{q}^{2}$. The gluon
propagators $G_{1,2}$, the charm propagators $P_{0},P_{1},P_{2}$ and every numerator $N_{X}$
and $N^{X}_{a,b}$ depend only on the hard scales $m_{c}$, $M$ and on the fixed kinematic
invariant $k\!\cdot\!K=(M^{2}-m^{2})/2$. Because $\hat q$ enters exclusively through
the charm-line momenta, this single light-quark propagator is inert under the spherical
$\hat q^{2}$ derivative and the identical factor $L$ appears in $I^{q}_{0}(u)$ and in
$I^{q}_{2}(u)$. The relativistic correction redistributes powers among the charm
propagators but introduces no new $u$- or $m_{q}$-dependent denominator. Whatever
insensitivity to $u$ and $m_{q}$ is exhibited by $I^{q}_{0}(u)$ is therefore inherited by
$I^{q}_{2}(u)$ at the structural level, independent of the more elaborate charm sector of
the latter, a fact that explains, at one stroke, both why the two kernels share the
same flatness pattern and why the property is not a leading-order accident.

This structural fact is turned into a quantitative insensitivity by the strong hierarchy
of scales that characterizes the decay,
\begin{equation}
m_{q}\,\lesssim\,0.1~\mathrm{GeV}\;\ll\;m\,\lesssim\,1~\mathrm{GeV}\;<\;m_{c}\simeq M/2.
\label{eq:hierarchy}
\end{equation}
The loop integral is controlled by the charm propagators, whose denominators place the
dominant support of the integrand at hard momenta $k_{1}^{2}\sim m_{c}^{2}$. Writing
$L=k_{1}^{2}-2u\,(k_{1}\!\cdot\!p)+u^{2}m^{2}-m_{q}^{2}$, the light-quark mass enters only as
$m_{q}^{2}$ and is turned by the loop integration into a chiral logarithm
$(m_{q}^{2}/m_{c}^{2})\ln(m_{c}^{2}/m_{q}^{2})$, which vanishes as $m_{q}\rightarrow0$ and stays
below 2\% for the physical value. The momentum fraction enters through the other
two terms. The linear term $2u\,(k_{1}\!\cdot\!p)$ is built from the hard product
$k_{1}\!\cdot\!p$ and is not small in itself, but it is odd about $u=1/2$ and cancels against its
$u\leftrightarrow\bar u$ partner in the symmetrized integrand of Eqs.~(\ref{eq:integq0})
and~(\ref{eq:integq2}), which enforces $I^{q}_{0,2}(u)=I^{q}_{0,2}(\bar u)$. The quadratic term
$u^{2}m^{2}$ carries only the soft scale $p^{2}=m^{2}$, so together with the symmetrization it
leaves in $L$ a residual $u\bar u\,m^{2}=[\tfrac14-(u-1/2)^{2}]\,m^{2}$. Its $u$-independent part
$\tfrac14 m^{2}$ merely shifts the denominator uniformly and produces no modulation, so the only
genuine $u$ dependence resides in the $(u-1/2)^{2}m^{2}$ piece. Set against the hard scale
$m_{c}^{2}\sim k_{1}^{2}$ that controls $L$, this soft shift changes the kernel only fractionally,
by an amount even about $u=1/2$ and of order $(u-1/2)^{2}m^{2}/m_{c}^{2}$, a few percent at most
across the physical range, and a direct evaluation of the scalar integrals confirms both this
estimate and the chiral-logarithmic dependence on $m_{q}$.

One might worry that this hard-region argument is undermined by the soft region of the loop,
where one gluon approaches its mass shell, $k_{1}\rightarrow p$, and the two gluon denominators
and the light-quark denominator vanish together. There an individual diagram is strongly enhanced
and does become sensitive to $m_{q}$ and to the shape of $\phi^{q}$, precisely the region in which
the soft scales could reassert themselves. This enhancement does not survive the sum over the six
diagrams. Their numerators combine into a single gauge-invariant structure that vanishes fast
enough as the soft configuration is approached to overcome the singularity of the propagators, so
that the soft region contributes negligibly and the summed kernel is infrared finite. The same
numerator zero cancels the light-quark pole when that line goes on shell, so that the chiral limit
$m_{q}\rightarrow0$ is reached smoothly. Gauge invariance thus switches off the one region that
could have reinstated a strong dependence on $u$ and $m_{q}$, and the kernel is left under the
control of the hard region alone.

This flatness is not a generic feature of hard exclusive amplitudes but depends on which line the
momentum fraction sits on. Here it sits on the soft, subdominant light-quark line, whose only
hard imprint is the linear term that the symmetrization removes. In the electromagnetic transition
form factor of the $\eta^{(\prime)}$, by contrast, the momentum fraction sits on the hard
active-quark propagator, whose virtuality is set by $u$ together with the large scale $Q^{2}$. Such
a hard propagator enters the amplitude as a pole $1/(uQ^{2})$, and the $u\leftrightarrow\bar u$
symmetrization adds the crossed term $1/(\bar uQ^{2})$. The
hard kernel is then the power $1/(u\bar u\,Q^{2})$, in which $u$ multiplies the hard scale rather
than a soft one, so that no scale hierarchy suppresses its variation and it grows without bound
toward the end points $u\rightarrow0,1$. The convolution accordingly weights the inverse moments of the DA and remains genuinely sensitive to the shape of the DA.

Physically, the near constancy of both kernels reflects the dominance of the hard charm
scale over every other scale in the process. The $c\bar c$ pair annihilates into a photon
and two virtual gluons whose typical virtualities are themselves of order $m_{c}^{2}$,
much larger than $m^{2}$ and $m_{q}^{2}$, so that the $g^{*}g^{*}\rightarrow\etapp$
subprocess is probed at a momentum scale well above any soft scale of the
$\eta^{(\prime)}$. To the hard subprocess the $\eta^{(\prime)}$ accordingly appears as a
quasipointlike composite object, much as a probe of large virtuality resolves only the
integrated structure of a hadron in the deep-inelastic regime. The flatness of the loop
kernel is the direct expression of this scale separation. Convolved with such a flat kernel,
the DA then contributes essentially through its normalization alone, its
zeroth Mellin moment, while its detailed shape and the value of $m_{q}$ enter only through the
small $(u-1/2)^{2}m^{2}/m_{c}^{2}$ modulation.

A further and independent sign that the effect is set by the hard charm scale, and not by the
particular kinematics of the radiative channel, comes from the closely related Dalitz process
$\Jpsi\rightarrow\etapp\,e^{+}e^{-}$ that we have studied in Ref.~\cite{He:2020jvj}. There the
real photon is replaced by a virtual one of invariant mass $q^{2}=M_{e^{+}e^{-}}^{2}$, which
introduces an additional external scale into the hard kernel. The same near independence of the
loop kernel on $u$ and $m_{q}$ is found to persist throughout the accessible dilepton region
$q^{2}\lesssim 1~\mathrm{GeV}^{2}$, for the simple reason that $m_{c}^{2}$ remains far larger than
$q^{2}$ and continues to fix the scale of the loop. The flatness is therefore a genuine property
of the hard charm loop, stable against a change of the external photon virtuality, and not an
artifact of the on-shell point $q^{2}=0$.

The flatness of the quark-content kernels carries a direct phenomenological consequence. Because
$I^{q}_{0}(u)$ and $I^{q}_{2}(u)$ vary so little across the integration region, the convolution that builds
the decay amplitude becomes insensitive to the detailed shape of the DA and to
the value of $m_{q}$ so that the amplitude is fixed almost entirely by the normalization of the
DA, namely the decay constants $f^{q}_{\etapp}$. The branching ratios
predicted for $\psi_{n}\rightarrow\gamma\etapp$ inherit this insensitivity. Section~\ref{sec:numerical}
demonstrates it directly with the numerical helicity amplitudes, which show how weakly the prediction
varies with the DA shape and with $m_{q}$.

\subsection{Gluon-content contribution}
\label{subsec:gluon}

\begin{figure}[H]
\centering
\includegraphics[width=0.5\textwidth]{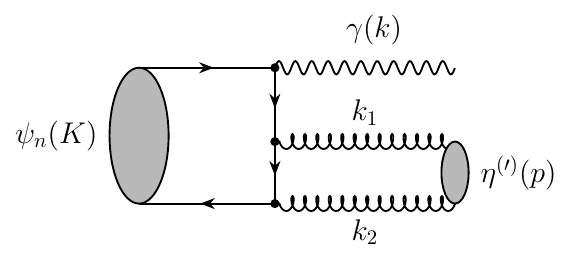}
\caption{A representative Feynman diagram for the gluon-content contribution to
$\psi_{n}\rightarrow\gamma\etapp$, in which the $c\bar c$ pair annihilates into a photon and two
collinear gluons that directly form the two-gluon content of the $\etapp$. Two further diagrams
are obtained by permuting the photon and gluon attachments on the charm line.}
\label{fig:gluon}
\end{figure}

The $\eta^{(\prime)}$ also carries an intrinsic two-gluon Fock component. The gluon-content
contribution proceeds through the same hard subprocess $\psi_{n}\rightarrow\gamma g^{*}g^{*}$, but
the two gluons now couple directly to this two-gluon component of the $\eta^{(\prime)}$
through its leading-twist gluonic DA, without the intermediate light-quark loop
of the quark contribution. The two gluons carry momentum fractions $u$ and $\bar u$ of the
$\eta^{(\prime)}$ momentum $p$, and since these are fixed by the DA the process
is tree level, with no loop integration. A representative diagram is shown in Fig.~\ref{fig:gluon}.

The leading-twist two-gluon matrix element is~\cite{Kroll:2002nt,Ball:2007hb,Agaev:2014wna}
\begin{equation}
\langle\etapp(p)|A^{a}_{\alpha}(x)A^{b}_{\beta}(y)|0\rangle
=\frac{1}{4}\epsilon_{\alpha\beta\rho\sigma}\frac{k^{\rho}p^{\sigma}}{p\cdot k}
\frac{C_{F}}{\sqrt 3}\frac{\delta^{ab}}{8}\,f^{g}_{\etapp}
\int_{0}^{1}\!du\,e^{i(up\cdot x+\bar u p\cdot y)}\frac{\phi^{g}(u)}{u\bar u},
\label{eq:gluonMT}
\end{equation}
with $f^{g}_{\etapp}=\tfrac{1}{\sqrt3}(f^{u}_{\etapp}+f^{d}_{\etapp}+f^{s}_{\etapp})$ and the gluonic
twist-2 DA
\begin{equation}
 \phi^{g}(u)=30\,u^{2}\bar u^{2}\!\!\sum_{n=2,4,\dots}\!\!c^{g}_{n}(\mu)C^{5/2}_{n-1}(2u-1).
 \label{eq:gluonDA}
\end{equation}
Contracting the $\psi_{n}\rightarrow\gamma g^{*}g^{*}$
amplitude of Eq.~(\ref{eq:assemble}) with this gluonic vertex, inserting the two gluon propagators at
the fixed momenta $up$ and $\bar u p$, and projecting onto the helicity amplitude with the
projector of Eq.~(\ref{eq:helproj}) yields, with the relativistic correction in the form of
Eq.~(\ref{eq:Aintegrated}),
\begin{equation}
H^{g}=\frac{2Q_{c}}{9}\sqrt{4\pi\alpha_{e}}\,(4\pi\alpha_{s})\,\frac{1}{\sqrt{\pi M}}
\frac{f^{g}_{\etapp}}{M}\int_{0}^{1}\!du\,\frac{\phi^{g}(u)}{u\bar u}\Big[\,R_{\psi_{n}}(0)\,I^{g}_{0}(u)-\nabla^{2}R_{\psi_{n}}(0)\,I^{g}_{2}(u)\,\Big].
\label{eq:Hg}
\end{equation}
The two kernels $I^{g}_{0}$ and $I^{g}_{2}$ are, as in the quark content, the value of the hard
amplitude at $\hat q=0$ and its $q^{2}$ correction, and the three diagrams are labeled
$X\in\{A,B,C\}$ by the position of the photon vertex on the charm line, as in the quark case. In the
weak-binding approximation $m_{c}=M/2$ the leading-order kernel reduces to the two terms of
topologies $A$ and $B$,
\begin{equation}
I^{g}_{0}(u)=\frac{N^{A}_{g}}{P_{0}\,P_{u}}+\frac{N^{B}_{g}}{P_{0}\,P_{\bar u}},
\label{eq:integg0}
\end{equation}
topology $C$, in which the photon is emitted between the two gluon vertices, being absent at this
order. Each diagram of that topology is proportional to the squared component of its gluon momentum
orthogonal to the plane spanned by the photon and quarkonium momenta. The leading-twist projection
assigns the gluons the collinear momenta $up$ and $\bar u p$ with vanishing transverse components,
and the contribution drops out. The three charm propagators on the line are
\begin{align}
P_{0}&=-k\!\cdot\!K+i\epsilon,\nonumber\\
P_{u}&=-u\big[M^{2}\bar u+(k\!\cdot\!K)(u-\bar u)\big]+i\epsilon,\nonumber\\
P_{\bar u}&=-\bar u\big[M^{2}u+(k\!\cdot\!K)(\bar u - u)\big]+i\epsilon,
\label{eq:gluondenom}
\end{align}
left in factored form rather than combined over the common denominator $P_{0}P_{u}P_{\bar u}$. The
propagator $P_{0}$ stands apart, since it carries no gluon momentum and is fixed by the external photon
and quarkonium momenta alone so that it is independent of $u$ and collapses to the constant
$-k\!\cdot\!K$, while $P_{u}$ and $P_{\bar u}$ carry the gluon momenta $up$ and $\bar u p$. Across the
physical interval the constant $P_{0}$ and the denominators $P_{u}$, $P_{\bar u}$ stay nonzero, the latter
reaching zero only at the end points $u=0,1$, where the gluonic DA suppresses the
integrand. No charm line is therefore brought on shell as $u$ is integrated, the hard kernel remains
real, and the gluon-content amplitude develops no absorptive part. This conclusion extends unchanged
to the $q^{2}$ correction, whose denominators are only higher powers $P_{0}^{a}P_{u}^{b}$, $P_{0}^{a}P_{\bar u}^{b}$, and $P_{u}^{a}P_{\bar u}^{b}$ of the same factors and open no new threshold. The numerators are
\begin{align}
N^{A}_{g}&=\tfrac{1}{2} (k\!\cdot\!K) M^2 u,
\label{eq:gluonN0}
\end{align}
with $N^{B}_{g}=-N^{A}_{g}(u\leftrightarrow \bar u)$. The relative minus sign reflects Bose
symmetry. Interchanging the two identical gluons swaps $u$ and $\bar u$ and maps topology $A$ into
$B$, while the antisymmetric $\epsilon_{\alpha\beta\rho\sigma}$ structure of the matrix
element of Eq.~(\ref{eq:gluonMT}) changes sign, so the kernel must be odd under $u\leftrightarrow\bar u$,
just as $\phi^{g}$ is. Summing the two terms reproduces the compact form of the leading-order gluon
kernel obtained in Ref.~\cite{He:2019},
\begin{equation}
I^{g}_{0}(u)=\frac{2x(2u-1)}{1-x^{2}(1-2u)^{2}},
\label{eq:gluonI0compact}
\end{equation}
where $x=m^{2}/M^{2}$ denotes the squared ratio of the $\etapp$ to the charmonium mass.

The $q^{2}$ correction is obtained by restoring the charm-line momenta as $K/2\pm\hat q$ and taking the
spherical average of the second derivative at $\hat q=0$, equivalent to the transverse projection
$\tfrac{1}{2}P_{T}^{\mu\nu}\partial^{2}/\partial\hat q^{\mu}\partial\hat q^{\nu}$.
Being tree level it involves no loop integration. The leading-order vanishing of topology $C$ rests
on the weak-binding configuration itself, with both charm quarks carrying the momentum $K/2$. The
relative momentum shifts the charm momenta to $K/2\pm\hat q$, the cancellation no longer operates
away from this symmetric point, and the second $\hat q$ derivative revives this topology, although
only through its lowest term. Collecting the result by the same powers $(a,b)$ as in Eq.~(\ref{eq:integq2}) gives
\begin{equation}
I^{g}_{2}(u)=\sum_{X\in\{A,B,C\}}\,\sum_{(a,b)}\,\frac{N^{X}_{g,a,b}}{D^{X}_{a,b}},
\label{eq:integg2}
\end{equation}
with $D^{A}_{a,b}=P_{0}^{a}P_{u}^{b}$, $D^{B}_{a,b}=P_{0}^{a}P_{\bar u}^{b}$, and $D^{C}_{a,b}=P_{u}^{a}P_{\bar u}^{b}$,
the sums for $X=A,B$ running over $a+b\le4$ and topology $C$ contributing only $(a,b)=(1,1)$, every
other term vanishing. The nonzero numerators of topology $A$ are
\begin{align}
N^{A}_{g,1,1}&=\tfrac{1}{6}\big(2(k\!\cdot\!K)+M^{2}\big), &
N^{A}_{g,1,2}&=\tfrac{1}{6}(k\!\cdot\!K)\,u\big(4(k\!\cdot\!K)\,u+3M^{2}\big),\nonumber\\
N^{A}_{g,2,1}&=\tfrac{1}{6}(k\!\cdot\!K)\,u\big(4(k\!\cdot\!K)+3M^{2}\big), &
N^{A}_{g,1,3}&=\tfrac{2}{3}(k\!\cdot\!K)^{3}u^{3},\nonumber\\
N^{A}_{g,2,2}&=\tfrac{2}{3}(k\!\cdot\!K)^{3}u^{2}, &
N^{A}_{g,3,1}&=\tfrac{2}{3}(k\!\cdot\!K)^{3}u,
\label{eq:gluonNA}
\end{align}
the topology-$C$ numerator $N^{C}_{g,1,1}=\tfrac{1}{3}(u-\bar u)\big((k\!\cdot\!K)-M^{2}\big)$, and
those of topology $B$ fixed by the symmetry $N^{B}_{g,a,b}=-N^{A}_{g,a,b}(u\leftrightarrow \bar u)$.

As the compact kernel of Eq.~(\ref{eq:gluonI0compact}) shows, at leading order the gluon contribution
is suppressed by the overall factor $x=m^{2}/M^{2}$, reflecting the near-on-shell gluons and the
Ore-Powell form of their coupling to the $\eta^{(\prime)}$~\cite{Krammer:1978qp,Billoire:1978xt}. The
$q^{2}$ correction is a relativistic effect smaller still, so the gluon contribution is subleading to the
quark one. Its dependence on the DA, on the other hand, differs in character from
the quark case. In the leading-order kernel the denominator departs from unity only at order $x^{2}$ and
varies little across the momentum-fraction range, much as in the quark case, while the numerator
carries an explicit and unsuppressed dependence on $u$. The gluon hard kernel is therefore genuinely
$u$ dependent rather than flat, and the convolution probes the shape of the gluonic DA rather than its normalization alone. The relativistic correction is built from the same
denominators raised to higher powers and from numerators that likewise carry the momentum-fraction
dependence, so the leading-order and $q^{2}$ gluon kernels probe the DA in the
same way. Both stand in contrast to the flatness established in Sec.~\ref{subsec:quark}, where the
hard charm scale dominating the loop integration strongly suppresses the momentum-fraction dependence. Moreover,
the gluon process involves no light-quark propagator, so unlike the quark contribution it carries no
dependence on the light-quark mass $m_{q}$.

\subsection{QED contribution}
\label{subsec:qed}

\begin{figure}[H]
\centering
\includegraphics[width=0.9\textwidth]{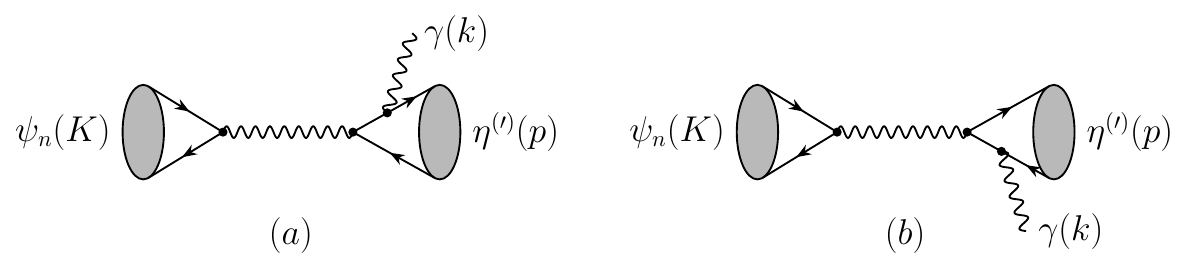}
\caption{Feynman diagrams for the QED contribution $\psi_{n}\rightarrow\gamma^{*}\rightarrow\gamma\etapp$,
in which the $c\bar c$ pair annihilates into a single virtual photon that produces the light
quark-antiquark pair of the $\etapp$. In contrast to the quark-antiquark and two-gluon
contributions, the real photon is radiated from a final-state light-quark line, the quark in (a)
and the antiquark in (b).}
\label{fig:qed}
\end{figure}

Finally, the $\eta^{(\prime)}$ can be produced electromagnetically. The $c\bar c$ pair annihilates into
a single virtual photon, $\psi_{n}\rightarrow\gamma^{*}$, which produces the light-quark--antiquark
pair of the $\eta^{(\prime)}$. The real photon is radiated from this light-quark line, leaving one
internal light-quark propagator, and the pair is projected onto the meson by the quark DA. This contribution shares the helicity structure and the bound-state projection of the quark
one. The two-gluon exchange between the charm and light-quark lines is replaced by a single
virtual-photon exchange, so this part of the amplitude is of order $\alpha_{e}$ instead of
$\alpha_{s}^{2}$ and carries the squared light-quark charges $Q_{q}^{2}$. The process is tree level and
is shown in Fig.~\ref{fig:qed}.

A distinctive feature of this contribution is that the $c\bar c$ pair annihilates at a single
$\gamma^{*}$ vertex, so the charm line carries no internal propagator. The hard amplitude is then
independent of the relative momentum $\hat q$, and the $q^{2}$ relativistic correction vanishes
identically, leaving the QED contribution purely at leading order,
\begin{equation}
H^{\rm QED}=-Q_{c}\,(4\pi\alpha_{e})^{3/2}\sqrt{\frac{3}{\pi M}}\,R_{\psi_{n}}(0)
\sum_{q=u,d,s}\frac{Q_{q}^{2}f^{q}_{\etapp}}{M}\int_{0}^{1}\!du\,\phi^{q}(u)\,I^{\rm QED}(u),
\label{eq:Hqed}
\end{equation}
with the $\hat q$-independent QED kernel
\begin{equation}
I^{\rm QED}(u)=\frac{1-x}{2}
\left(\frac{1}{u-u\bar u\,x-m_{q}^{2}/M^{2}+i\epsilon}+(u\leftrightarrow\bar u)\right).
\label{eq:IQED}
\end{equation}
This kernel fixes how the QED amplitude depends on the light-quark mass and on the DA. The light-quark mass appears only through the strongly power-suppressed ratio $m_{q}^{2}/M^{2}$ in the
propagator, so that the kernel is almost independent of it. With the
$i\epsilon$ prescription the propagator acquires an imaginary part only where its denominator vanishes,
at $u\simeq m_{q}^{2}/M^{2}$. This point lies very close to the end point $u=0$, where the DA nearly vanishes, so the absorptive part is very small and scales as $m_{q}^{2}/M^{2}$.

The two terms of Eq.~(\ref{eq:IQED}) form a $u\leftrightarrow\bar u$ pair, so the kernel is even under
this exchange. It is peaked near the end points rather than flat, but its $1/(u\bar u)$ growth is
offset by the end point suppression $u\bar u$ of the leading-twist DA, leaving a
regular convolution. This evenness leaves the convolution dominated by the DA normalization, and the
shape enters only through the low Gegenbauer moments. The QED amplitude is therefore set mainly by the decay constants and is weakly
sensitive to the shape. This is opposite
to that of the gluon kernel of Eq.~(\ref{eq:gluonI0compact}), which
is odd in $u\leftrightarrow\bar u$ and so weights the shape of the DA rather than
its normalization, and it is distinct from the quark contribution, where the flatness produced by the
loop integration removes the shape dependence directly.

\section{Numerical analysis}
\label{sec:numerical}

\subsection{Decay width and input parameters}
\label{subsec:inputs}

The partial width for $\psi_{n}\rightarrow\gamma\etapp$ is fixed by the modulus squared of
the single helicity amplitude, to which the quark, gluon, and QED contributions add coherently,
\begin{equation}
\Gamma(\psi_{n}\rightarrow\gamma\etapp)
=\frac{M^{2}-m^{2}}{16\pi M^{3}}\frac{2}{3}\,\big|H^{q}+H^{g}+H^{\rm QED}\big|^{2}.
\label{eq:width}
\end{equation}
The first factor is the two-body phase space for a massless photon recoiling against the
$\etapp$. With the quantization axis taken along the $\etapp$ flight
direction,\footnote{We choose $+\hat z$ along the $\etapp$ momentum, so that the photon
moves along $-\hat z$. This convention fixes the nonzero helicity configurations to coincide with
those of Ref.~\cite{Korner:1982vg} and underlies the helicity-projector construction of
Sec.~\ref{sec:framework}.} conservation of the angular momentum along the axis,
$\lambda_{V}=\lambda_{\eta}-\lambda_{\gamma}=-\lambda_{\gamma}$, permits only the two helicity
configurations $(\lambda_{V},\lambda_{\gamma})=(+1,-1)$ and $(-1,+1)$, which are equal by parity and
define the single independent amplitude $H$. The factor $2$ is then the sum over these two nonzero
terms, and the factor $1/3$ the average over the three spin states of the initial $\psi_{n}$.

Of the three contributions the quark term, in which the $c\bar c$ pair annihilates into
$\gamma g^{*}g^{*}$ and the two gluons materialize the $\etapp$ through its leading
quark-antiquark Fock component, dominates both channels. The gluon contribution, in which the two
collinear gluons couple to the two-gluon component of the $\etapp$, enters only as a
correction. As shown in Sec.~\ref{subsec:gluon}, its hard kernel scales with the ratio $m^{2}/M^{2}$
of the light meson to the charmonium mass and vanishes in the chiral limit, so the gluon contribution
is strongly suppressed for the lighter $\eta$ and grows for the heavier and more singletlike
$\etap$. The QED contribution $c\bar c\rightarrow\gamma^{*}\rightarrow\gamma\etapp$,
in which the light quarks of the $\etapp$ are produced electromagnetically rather than
through the two gluons, is suppressed relative to the quark amplitude by $\sim\alpha_{e}/\alpha_{s}^{2}$
and by the light-quark electric charges, and is the smallest. The quark and gluon amplitudes are
complex while the QED amplitude is real, the absorptive parts arising from internal propagators that
reach their mass shell.

The evaluation of Eq.~(\ref{eq:width}) requires three groups of inputs. These are the charmonium and
meson kinematics together with the nonrelativistic wave-function parameters, the DAs of the quark and gluon Fock components of the $\etapp$, and the
$\eta$--$\etap$ mixing parameters that fix the flavor decay constants. We specify them in turn. The
charmonium masses $M_{J/\psi}=3.097$ and $M_{\psi(2S)}=3.686~\mathrm{GeV}$, the meson
masses $m_{\eta}=0.548$ and $m_{\etap}=0.958~\mathrm{GeV}$, the fine-structure constant $\alpha_{e}=1/137$, and the strong
coupling $\alpha_{s}(M_{Z})=0.1180$ are taken from the Particle Data Group~\cite{ParticleDataGroup:2024}. The strong
coupling is run at one loop to the hard scale $\mu=M_{\psi_{n}}/2$ of each decay, giving
$\alpha_{s}(M_{J/\psi}/2)=0.294$ and $\alpha_{s}(M_{\psi(2S)}/2)=0.276$.

The first nonperturbative ingredient is the charmonium wave function at the origin. The leading-order
amplitude is governed by the wave function itself, $R_{\psi_{n}}(0)$, and the relativistic correction by
its Laplacian $\nabla^{2}R_{\psi_{n}}(0)$, which is controlled by the internal motion of the charm
quarks. Both are fixed from the Cornell potential of Eichten and Quigg in its
frozen-$\alpha_{s}$ form, a Coulomb-plus-linear potential whose coupling runs at short distance and
saturates at long distance~\cite{Eichten:2019gig,Eichten:1978tg}. The wave functions at the origin,
$|R_{J/\psi}(0)|^{2}=1.0952$ and $|R_{\psi(2S)}(0)|^{2}=0.6966~\mathrm{GeV}^{3}$, are taken from the
tabulation of Eichten and Quigg~\cite{Eichten:2019hbb}, and solving the radial Schr\"odinger equation
in the same potential gives the ratios
\begin{equation}
\frac{\nabla^{2}R_{J/\psi}(0)}{R_{J/\psi}(0)}=-0.53~\mathrm{GeV}^{2},\qquad
\frac{\nabla^{2}R_{\psi(2S)}(0)}{R_{\psi(2S)}(0)}=-1.59~\mathrm{GeV}^{2}.
\label{eq:lapR}
\end{equation}
The $\psi(2S)$ value is 3 times the $J/\psi$ one, the radially excited state being spatially
more extended and less tightly bound, so that the $q^{2}$ corrections are intrinsically far more
important there. For the $J/\psi$ the ratio in Eq.~(\ref{eq:lapR}) agrees with the Cornell-potential
evaluations of the corresponding order-$v^{2}$ NRQCD matrix element in
Refs.~\cite{Bodwin:2006dn,Bodwin:2007fm}, and
for the $\psi(2S)$ it is somewhat larger than the value adopted in Ref.~\cite{Kivel:2023rp}, which
follows from the Gremm-Kapustin relation with the binding energy taken as $M_{\psi(2S)}-2m_{c}$.
The wave function at the origin itself can be checked independently, since the leptonic width
$\Gamma(\psi_{n}\to e^{+}e^{-})$ is fixed by $R_{\psi_{n}}(0)$. Evaluated from the leptonic-width formula with its first-order QCD
correction~\cite{Mackenzie:1981sf}, using the $\alpha_{s}$ values above, the adopted
$R_{\psi_{n}}(0)$ reproduce the measured $\Gamma(J/\psi\to e^{+}e^{-})$ and
$\Gamma(\psi(2S)\to e^{+}e^{-})$~\cite{ParticleDataGroup:2024} to about $2\%$ and $10\%$, respectively.
This check is what controls the potential-model dependence of the predictions, which would otherwise
be substantial. Among the four standard potentials tabulated earlier by the same
authors~\cite{Eichten:1995ch}, $|R_{1S}(0)|^{2}$ ranges from $0.810$ to $1.454~\mathrm{GeV}^{3}$ and
$|R_{2S}(0)|^{2}$ from $0.418$ to $0.927~\mathrm{GeV}^{3}$, ranges that cover the values adopted here,
and their constituent charm masses lie between $1.48$ and $1.84~\mathrm{GeV}$, a range that likewise
covers the weak-binding value $m_{c}=M/2$ used in the hard kernel. Because the radiative and the
leptonic widths carry the same factor $|\Rzero|^{2}$, it cancels in their ratio, so that the branching
ratios given below are in effect predictions of
$\Gamma(\psi_{n}\to\gamma\etapp)/\Gamma(\psi_{n}\to e^{+}e^{-})$, with the measured leptonic width
fixing their overall size. What remains of the model dependence is then the accuracy with which the
leptonic widths are reproduced, 2\% and 10\%. Beyond that common factor, the relativistic
correction involves the wave function only through the ratio $\nabla^{2}\Rzero/\Rzero$, which was
compared with other determinations above.

The DAs are the second nonperturbative ingredient. With the
conventions of Sec.~\ref{sec:framework}, the leading-twist quark-antiquark DA is fixed by its first
two Gegenbauer moments $a^{q}_{2}$ and $a^{q}_{4}$ and the two-gluon DA by its lowest moment
$a^{g}_{2}$. Because the $\etapp$ amplitudes are not precisely known, we use the three representative
models of Table~\ref{tab:lcda}, taken from the $\etapp$ transition-form-factor analysis of
Ref.~\cite{Agaev:2014wna}. At the reference scale $\mu_{0}=1~\mathrm{GeV}$ the three profiles differ markedly. Model~I
has small moments and is single humped, close to the asymptotic amplitude $6u\bar u$ of perturbative
QCD~\cite{Lepage:1980fj,Efremov:1979qk}, and a single broad peak centered at $u=1/2$.
Model~II, with $a^{q}_{2}=0.20$ and $a^{q}_{4}=0$, is broader and flat topped, its central value
lowered below the asymptotic one, of the moderate type favored by light-cone sum
rules~\cite{Ball:2007hb}. Model~III, with the largest $a^{q}_{2}=0.25$ and a negative
$a^{q}_{4}=-0.10$, suppresses the midpoint into a double-humped profile, with a dip at $u=1/2$
and two maxima displaced toward the end points, in the manner of the Chernyak-Zhitnitsky
amplitude~\cite{Chernyak:1981zz,Chernyak:1983ej}. The $a^{q}_{2}$ values of the three models span
those of the more recent lattice-QCD~\cite{RQCD:2019osh,Baker:2024zcd} and QCD
sum-rule~\cite{Zhong:2022lmn} determinations, so the set still brackets the present spread of
$\etapp$ DAs. The moments are quoted at $\mu_{0}=1~\mathrm{GeV}$ and evolved to the hard scale
$\mu=M/2$ by the coupled quark-gluon ERBL equations~\cite{Kroll:2002nt}, which damp the
differences as the moments shrink toward the asymptotic limit.

\begin{table}[H]
\caption{The three DA models of Ref.~\cite{He:2019}. The table lists
the Gegenbauer moments $a^{q}_{2}$, $a^{q}_{4}$ of the quark DA and $a^{g}_{2}$ of the two-gluon DA at
$\mu_{0}=1~\mathrm{GeV}$. The spread among the models quantifies the DA systematic.}
\label{tab:lcda}
\centering
\begin{ruledtabular}
\begin{tabular}{l D{.}{.}{2} D{.}{.}{2} D{.}{.}{2}}
Model & \multicolumn{1}{c}{$a^{q}_{2}$} & \multicolumn{1}{c}{$a^{q}_{4}$} & \multicolumn{1}{c}{$a^{g}_{2}$} \\ \hline
I   & 0.10 & 0.10  & -0.26 \\
II  & 0.20 & 0.00  & -0.31 \\
III & 0.25 & -0.10 & -0.25 \\
\end{tabular}
\end{ruledtabular}
\end{table}

The third ingredient is the $\eta$--$\etap$ mixing. In the quark-flavor basis the physical $\eta$
and $\etap$ are superpositions of the unphysical states $\eta_{q}=(u\bar u+d\bar d)/\sqrt2$ and
$\eta_{s}=s\bar s$, rotated into the mass eigenstates by a single angle $\phi$,
\begin{equation}
\begin{pmatrix}\eta\\[2pt]\etap\end{pmatrix}
=\begin{pmatrix}\cos\phi & -\sin\phi\\[2pt]\sin\phi & \cos\phi\end{pmatrix}
\begin{pmatrix}\eta_{q}\\[2pt]\eta_{s}\end{pmatrix}.
\label{eq:mixangle}
\end{equation}
In the FKS scheme~\cite{Feldmann:1998vh}, where the
decay constants $f_{q}$ and $f_{s}$ of the $\eta_{q}$ and $\eta_{s}$ components are assumed to be
common to the $\eta$ and the $\etap$, their flavor decay constants are fixed by these two constants
and the single mixing angle $\phi$,
\begin{equation}
\begin{aligned}
f^{u(d)}_{\eta}&=\frac{f_{q}}{\sqrt2}\cos\phi, & \qquad f^{s}_{\eta}&=-f_{s}\sin\phi,\\
f^{u(d)}_{\etap}&=\frac{f_{q}}{\sqrt2}\sin\phi, & \qquad f^{s}_{\etap}&=f_{s}\cos\phi.
\end{aligned}
\label{eq:fks}
\end{equation}
These three parameters are phenomenological, fixed not from first principles but from physical
processes so that different determinations return somewhat different values, the mixing angle in
particular varying by several degrees. We therefore take two representative determinations as
inputs, collected in Table~\ref{tab:mixing}. They agree on $f_{q}$ but differ appreciably in the
mixing angle. The lattice calculation of Ref.~\cite{ETMC:2025} gives the larger
angle, $\phi\simeq39^{\circ}$, which sits close to the long-standing value of the FKS analysis of
low-energy $\eta$--$\etap$ phenomenology~\cite{Feldmann:1998vh}. The transition-form-factor (TFF) analysis
of Ref.~\cite{Escribano:2013kba} gives a markedly smaller angle, $\phi\simeq33.5^{\circ}$, of the
kind preferred by hard, short-distance amplitudes, as we also found at leading order from
$J/\psi\rightarrow\gamma\etapp$ itself~\cite{He:2019}. We evaluate the $J/\psi$ observables in both,
adopting the lattice scheme as the default and retaining the $\etap$TFF scheme for the comparison of
total widths.

\begin{table}[H]
\caption{The two $\eta$--$\etap$ mixing schemes in the quark-flavor basis~\cite{Feldmann:1998vh},
given by the decay constants $f_{q}$, $f_{s}$ and the mixing angle $\phi$. ``Lattice'' denotes the determination of the Extended Twisted Mass
Collaboration~\cite{ETMC:2025} and ``$\etap$TFF'' the transition-form-factor
analysis~\cite{Escribano:2013kba}, the latter evaluated with $f_{\pi}=0.1302$~GeV.}
\label{tab:mixing}
\centering
\begin{ruledtabular}
\begin{tabular}{lccc}
Scheme & $f_{q}$~(GeV) & $f_{s}$~(GeV) & $\phi$~($^{\circ}$) \\ \hline
Lattice       & 0.1386(44) & 0.1707(33) & 39.3(2.0) \\
$\etap$TFF    & 0.1419(26) & 0.1250(52) & 33.5(0.9) \\
\end{tabular}
\end{ruledtabular}
\end{table}

\subsection{Results for the $J/\psi$}
\label{subsec:Jpsi}

We now present the numerical results, taking $J/\psi\to\gamma\etapp$ as the example and working in
the lattice mixing scheme, the $\etap$TFF scheme yielding the same pattern. The purpose of this
subsection is to exhibit how the helicity amplitudes depend on the two soft inputs of the hard
mechanism, the DA and the light-quark mass $m_{q}$. This near insensitivity
was already noted at leading order in our earlier study~\cite{He:2019}. Here we confirm it and
examine how the $q^{2}$ correction depends on the same two inputs. We scan the three DA models of Table~\ref{tab:lcda}, evolved to
the hard scale $\mu=M_{\psi_{n}}/2$, and three light-quark masses $m_{q}=0.01,0.05,0.10~\mathrm{GeV}$
whose range brackets the physical $u$, $d$, and $s$ masses.

The quark, gluon, and QED amplitudes are collected in Tables~\ref{tab:ampq}--\ref{tab:ampqed}. The quark and QED tables list nine entries each, for the three DA models and three
light-quark masses, and the gluon table three, one per DA model. For the quark and gluon components we
separate the helicity amplitude into a leading-order part $H^{X}_{\rm LO}$ and a $q^{2}$ relativistic
correction $H^{X}_{q^{2}}$, with $H^{X}=H^{X}_{\rm LO}+H^{X}_{q^{2}}$ ($X=q,g$) so that the size of the
correction can be read off directly. The QED component receives no $q^{2}$ relativistic correction,
with its helicity amplitude denoted $H^{\rm QED}_{\etapp}$. All amplitudes are in units of
$10^{-3}~\mathrm{GeV}$. Of the three contributions the quark amplitude is by far the largest. The
gluon amplitude is about $4\%$ of it for $\gamma\eta$ and $11\%$ for $\gamma\etap$, and the QED
amplitude about $24\%$ for $\gamma\eta$ and $7\%$ for $\gamma\etap$. The three add constructively but differ in analytic structure. The quark amplitude is a genuine one-loop object and is complex, its imaginary,
absorptive part generated wherever the internal light-quark and gluon lines of the
$g^{*}g^{*}\rightarrow\etapp$ subgraph reach their mass shell over the bulk of the momentum-fraction
integration. The QED and gluon amplitudes are real. In the QED term the single light-quark propagator reaches its
mass shell only near the end point of the momentum fraction, where the quark DA
vanishes, so its absorptive part is negligible.\footnote{The QED amplitude
carries a negligible imaginary part, about $10^{-9}~\mathrm{GeV}$ at the representative
$m_{q}=0.01~\mathrm{GeV}$ against a real part of order $10^{-4}~\mathrm{GeV}$; it comes from the
light-quark pole of the kernel near the end point $u\simeq m_{q}^{2}/M^{2}$, is suppressed by
$m_{q}^{2}/M^{2}$, and is discarded.} The gluon amplitude comes from $c\bar c$ annihilation into two
collinear gluons, and in the weak-binding kinematics $m_{c}=M/2$ its two charm propagators
never reach their mass shell, so it develops no absorptive part either. The total is thus quark dominated, the small real gluon and QED
terms adding to its real part while its imaginary part is carried by the quark loop alone.

\begin{table}[H]
\caption{Quark-component helicity amplitude for $J/\psi\to\gamma\etapp$ in the lattice mixing
scheme, in units of $10^{-3}~\mathrm{GeV}$, for the three DA models and
$m_{q}=0.01,0.05,0.10~\mathrm{GeV}$. $H^{q}_{\rm LO}$ is the leading order and $H^{q}_{q^{2}}$ the
$q^{2}$ correction, with $H^{q}=H^{q}_{\rm LO}+H^{q}_{q^{2}}$.}
\label{tab:ampq}
\centering
\begin{ruledtabular}
\begin{tabular}{l c c c c c}
 & & \multicolumn{2}{c}{$J/\psi\to\gamma\eta$} & \multicolumn{2}{c}{$J/\psi\to\gamma\etap$} \\
\cline{3-4}\cline{5-6}
Model & $m_{q}$ (GeV) & $H^{q}_{\rm LO}$ & $H^{q}_{q^{2}}$ & $H^{q}_{\rm LO}$ & $H^{q}_{q^{2}}$ \\ \hline
I & 0.01 & $-1.01-0.31i$ & $-0.45-0.07i$ & $-5.31-3.58i$ & $-1.86-1.68i$ \\
I & 0.05 & $-1.02-0.30i$ & $-0.45-0.06i$ & $-5.32-3.57i$ & $-1.87-1.66i$ \\
I & 0.10 & $-1.02-0.29i$ & $-0.44-0.04i$ & $-5.36-3.53i$ & $-1.90-1.62i$ \\
II & 0.01 & $-1.02-0.31i$ & $-0.46-0.08i$ & $-5.31-3.61i$ & $-1.87-1.71i$ \\
II & 0.05 & $-1.02-0.31i$ & $-0.45-0.06i$ & $-5.33-3.60i$ & $-1.88-1.69i$ \\
II & 0.10 & $-1.02-0.29i$ & $-0.44-0.04i$ & $-5.37-3.56i$ & $-1.91-1.65i$ \\
III & 0.01 & $-1.02-0.31i$ & $-0.46-0.07i$ & $-5.32-3.63i$ & $-1.88-1.72i$ \\
III & 0.05 & $-1.02-0.31i$ & $-0.46-0.06i$ & $-5.33-3.62i$ & $-1.89-1.70i$ \\
III & 0.10 & $-1.03-0.29i$ & $-0.44-0.04i$ & $-5.38-3.58i$ & $-1.91-1.66i$ \\
\end{tabular}
\end{ruledtabular}
\end{table}

\begin{table}[H]
\centering
\caption{Gluon-component helicity amplitude for $J/\psi\to\gamma\etapp$ in the lattice mixing
scheme, in units of $10^{-3}~\mathrm{GeV}$, with $H^{g}_{\rm LO}$ the leading order and $H^{g}_{q^{2}}$
the $q^{2}$ correction. The component is independent of $m_{q}$.}
\label{tab:ampg}
\begin{ruledtabular}
\begin{tabular}{l c c c c}
 & \multicolumn{2}{c}{$J/\psi\to\gamma\eta$} & \multicolumn{2}{c}{$J/\psi\to\gamma\etap$} \\
\cline{2-3}\cline{4-5}
Model & $H^{g}_{\rm LO}$ & $H^{g}_{q^{2}}$ & $H^{g}_{\rm LO}$ & $H^{g}_{q^{2}}$ \\ \hline
I   & $-0.03$ & $-0.03$ & $-0.51$ & $-0.50$ \\
II  & $-0.03$ & $-0.03$ & $-0.61$ & $-0.60$ \\
III & $-0.03$ & $-0.03$ & $-0.49$ & $-0.48$ \\
\end{tabular}
\end{ruledtabular}
\end{table}

\begin{table}[H]
\centering
\caption{QED-component helicity amplitude for $J/\psi\to\gamma\etapp$ in the lattice mixing
scheme, in units of $10^{-3}~\mathrm{GeV}$, for the three DA models and
$m_{q}=0.01,0.05,0.10~\mathrm{GeV}$. The tabulated $H^{\rm QED}$ is real and
receives no $q^{2}$ correction.}
\label{tab:ampqed}
\begin{ruledtabular}
\begin{tabular}{l c c c}
Model & $m_{q}$ (GeV) & $H^{\rm QED}_{\eta}$ & $H^{\rm QED}_{\etap}$ \\ \hline
I   & 0.01 & $-0.36$ & $-0.58$ \\
I   & 0.05 & $-0.36$ & $-0.58$ \\
I   & 0.10 & $-0.37$ & $-0.59$ \\
II  & 0.01 & $-0.36$ & $-0.58$ \\
II  & 0.05 & $-0.36$ & $-0.58$ \\
II  & 0.10 & $-0.37$ & $-0.59$ \\
III & 0.01 & $-0.35$ & $-0.56$ \\
III & 0.05 & $-0.35$ & $-0.56$ \\
III & 0.10 & $-0.35$ & $-0.57$ \\
\end{tabular}
\end{ruledtabular}
\end{table}

As Table~\ref{tab:ampq} shows, the dominant quark amplitude is almost completely insensitive to the two soft inputs of the hard
mechanism, the DA and the light-quark mass. Across
$m_{q}=0.01$--$0.10~\mathrm{GeV}$ it moves by less than $1\%$, and across
the three DA models it varies by less than $2\%$, so the leading contribution is fixed to high
accuracy independent of these poorly constrained quantities. This robustness was already
established at leading order in our earlier work~\cite{He:2019} and is now seen to survive the
$q^{2}$ correction. The QED amplitude in Table~\ref{tab:ampqed} is equally insensitive, changing by about $2\%$
across the $m_{q}$ range and by about $3\%$ across the DA models. The gluon amplitude carries no
light-quark propagator and is accordingly independent of $m_{q}$, but it is the one component with a
non-negligible dependence on the DA, as Table~\ref{tab:ampg} shows. Notably, the two gluon amplitudes depend on the DA
by a similar amount, but the $\gamma\eta$ one is so small that its variation falls below
$10^{-5}~\mathrm{GeV}$ and it appears unchanged in the table. The quark and QED kernels are smooth
functions of the momentum fraction whose convolution washes out the higher Gegenbauer moments, so
these two barely respond to the DA shape. The gluon instead couples directly to the two-gluon
DA and its kernel preserves that shape, so the model-to-model change of the
gluon moment passes almost undiluted into the amplitude, which varies by up to about $20\%$.
This is the largest variation among the three component amplitudes, yet the gluon amplitude is itself
the smallest of the three, so in the coherent sum the variation amounts to only about $1\%$,
leaving the full amplitude essentially independent of the DA. We note in passing that these entries carry no mixing-parameter error, since the amplitudes barely
depend on the light-quark mass and are thus blind to its flavor, so the mixing factors out as a single
overall weight, independent of the DA and $m_{q}$, that leaves the dependence shown
here untouched. We therefore defer its uncertainty to the branching ratios below rather than repeat it
on each amplitude.

The relativistic correction acts differently on the three contributions. The quark term, which
dominates, is enhanced substantially, its modulus $|H^{q}|$ growing by about $40\%$ in both channels
while its phase shifts by only a couple of degrees, so the correction scales the dominant amplitude up
rather than rotating it. The absorptive part of $H^{q}$ comes from the internal light-quark and gluon lines of the
loop going on shell, and it shrinks with the meson mass, vanishing as $x=m^{2}/M^{2}\to0$. At leading
order this reproduces the behavior found in our earlier study~\cite{He:2019}, and the $q^{2}$
correction, built from higher powers of these propagators, has an absorptive part that vanishes faster
still. The gluon term, though small, is enhanced
most strongly in relative terms, its $q^{2}$ piece comparable to the leading order and roughly
doubling the real gluon amplitude. The QED term receives no correction, its single-point $c\bar c$
annihilation leaving no charmonium relative momentum in the kernel. Summed over the three, the
correction enlarges the total $|H|$ by about a third to $40\%$. Its size, comparable to the
charmonium mean squared velocity $\langle v^{2}\rangle\simeq0.3$, is what one expects of the leading
relativistic correction in the velocity expansion of charmonium~\cite{Bodwin:1994jh,Bodwin:2006dn,Brambilla:2004wf,Brambilla:2010cs}.
In an analysis carried to the same order the $q^{2}$ correction to the three-gluon width of the
$\Jpsi$ is $77\%$~\cite{He:2025Vggg}, comparable in size to the $70\%$--$80\%$ by which the widths
grow here.

The radiative widths follow from the quark, gluon, and QED amplitudes through Eq.~\eqref{eq:width} as
the modulus squared of their coherent sum weighted by the two-body phase space. This modulus squared
is expanded and truncated at order $q^{2}$, so that the relativistic correction enters through its
interference with the leading order. The near insensitivity to the soft inputs carries over to the
widths, which move by only about $5\%$ across the three DA models and the $m_{q}$ range, so we
adopt the representative model~I at $m_{q}=0.01~\mathrm{GeV}$ and let the mixing parameters carry the
theoretical uncertainty. A branching ratio is what an experiment measures most directly, so we present our results in that
form and compare them with the measured values. Table~\ref{tab:widthsJpsi} gives the leading-order
branching ratio $\mathcal{B}_{\rm LO}$ and the $q^{2}$-corrected one $\mathcal{B}_{q^{2}}$ in the two
mixing schemes, together with the measured $\mathcal{B}_{\rm exp}$ from the Particle Data
Group~\cite{ParticleDataGroup:2024}, the error quoted for each scheme combining that scheme's mixing
parameters with the measured total width. The leading order typically falls short of the data by a factor
of two to three, and the relativistic correction is essential in closing the gap, increasing each
branching ratio by roughly a factor of $1.8$. With it the
$\etap$TFF scheme reaches about $55\%$ of the measured $\gamma\etap$ branching ratio and about $60\%$
of the $\gamma\eta$ one, while the lattice scheme reproduces the $\gamma\etap$ branching ratio to
within $25\%$ but undershoots $\gamma\eta$ by a factor of about 7, an imbalance that originates
in the mixing angle and is taken up below.

The remaining deficit may be attributed to the
uncertainties in $R_{\psi_{n}}(0)$ and $\alpha_{s}(\mu)$, since the dominant quark contribution
scales as $|R_{\psi_{n}}(0)|^{2}\alpha_{s}^{4}$. Neither is included in the quoted errors. The value of
$|R_{J/\psi}(0)|^{2}$ varies by nearly a factor of 2 among the potential models quoted in
Sec.~\ref{subsec:inputs}, which would carry over directly into the branching ratios, but the one
adopted here reproduces $\Gamma(J/\psi\to e^{+}e^{-})$ to $2\%$, and it is that agreement, not that
variation, that measures how well the leading-order amplitude is fixed. What survives enters through
the relativistic correction, a variation of $\nabla^{2}R_{J/\psi}(0)/R_{J/\psi}(0)$ by $\pm30\%$
moving the branching ratios by $\pm13\%$. The strong coupling is particularly
delicate at the charm scale, where its slowly converging running renders the one-loop value
$\alpha_{s}(M_{J/\psi}/2)=0.294$ adopted here appreciably lower than the two-loop value
$\alpha_{s}(M_{J/\psi}/2)=0.34$ used in Ref.~\cite{He:2019}. Replacing the one-loop value by the
two-loop one raises the branching ratios by $60\%$--$70\%$, while moving the scale from $\mu=M/2$ to
$\mu=M$ at one loop lowers them to between $40\%$ and $50\%$ of their central values, so that they are
determined only up to a factor of about two either way. This sensitivity is intrinsic to a hard
kernel of leading order in $\alpha_{s}$, where the $\mu$ dependence carried by $\alpha_{s}^{4}(\mu)$ is
left uncompensated. It is compensated only against the logarithms of the next order, so that a
next-to-leading-order calculation, a two-loop computation here, would be expected to narrow the
range. It is worth noting that with the two-loop coupling the $\etap$TFF results rise to
$\mathcal{B}_{q^{2}}(\gamma\eta)=1.10(11)\times10^{-3}$ and
$\mathcal{B}_{q^{2}}(\gamma\etap)=4.82(22)\times10^{-3}$, in very good agreement with the measured
values, and that the leading-order branching ratios are consistent with those of
Ref.~\cite{He:2019}.

\begin{table}[H]
\caption{Branching ratios $\mathcal{B}(J/\psi\rightarrow\gamma\etapp)$ in units of $10^{-3}$, for the
representative model~I DA at $m_{q}=0.01~\mathrm{GeV}$, in the lattice and
$\etap$TFF mixing schemes. Here $\mathcal{B}_{\rm LO}$ is the leading-order branching ratio,
$\mathcal{B}_{q^{2}}$ the one with the $q^{2}$ relativistic correction included, and
$\mathcal{B}_{\rm exp}$ the measured value, while the last row gives the ratio $\mathcal{R}_{1S}$ of
the $\gamma\etap$ to the $\gamma\eta$ branching ratio.}
\label{tab:widthsJpsi}
\centering
\begin{ruledtabular}
\begin{tabular}{l cc cc c}
 & \multicolumn{2}{c}{Lattice} & \multicolumn{2}{c}{$\etap$TFF} & \\
\cline{2-3}\cline{4-5}
Process & $\mathcal{B}_{\rm LO}$ & $\mathcal{B}_{q^{2}}$ & $\mathcal{B}_{\rm LO}$ & $\mathcal{B}_{q^{2}}$ & $\mathcal{B}_{\rm exp}$~\cite{ParticleDataGroup:2024} \\ \hline
$J/\psi\rightarrow\gamma\eta$  & $0.09(4)$ & $0.15(6)$  & $0.38(4)$ & $0.66(7)$  & $1.090(13)$ \\
$J/\psi\rightarrow\gamma\etap$ & $2.25(10)$ & $4.01(17)$ & $1.59(8)$ & $2.83(14)$ & $5.28(6)$ \\
$\mathcal{R}_{1S}$ & $24(10)$ & $26(11)$ & $4.2(5)$ & $4.3(6)$ & $4.84(8)$ \\
\end{tabular}
\end{ruledtabular}
\end{table}

The two channels pull in opposite directions on the individual branching ratios, so the sharper test is
their ratio $\mathcal{R}_{1S}=\mathcal{B}(J/\psi \to \gamma\etap)/\mathcal{B}(J/\psi \to \gamma\eta)$.
In it the dependence on the wave function at the origin $R_{J/\psi}(0)$ and the coupling $\alpha_{s}$
largely cancels. As these inputs are the main source of model and scale dependence,
$\mathcal{R}_{1S}$ is a far more reliable prediction than either branching ratio alone. Both channels are built from the
same mixing parameters $f_{q}$, $f_{s}$, and $\phi$, so in propagating their uncertainty to
$\mathcal{R}_{1S}$ we keep numerator and denominator correlated. The last row of
Table~\ref{tab:widthsJpsi} compares the two schemes with the measured
$\mathcal{R}_{1S}=4.84(8)$. The $\etap$TFF scheme is consistent with it, $\mathcal{R}_{1S}=4.2(5)$ at
leading order and $4.3(6)$ with the $q^{2}$ correction, close to, though slightly below, the measured value,
whereas the lattice scheme gives $\mathcal{R}_{1S}=24(10)$ and $26(11)$, far too large even allowing
for its inflated uncertainty. The difference is
the mixing angle. The larger lattice angle drives the $\eta$ flavor combination
$\sqrt2 f_{q}\cos\phi-f_{s}\sin\phi$ toward its zero, which suppresses the $\gamma\eta$ rate and
inflates the ratio. Because the combination is small there, the $\gamma\eta$ channel responds sharply to
$\phi$, so the mixing-angle uncertainty dominates the error and swells that of the lattice
$\mathcal{R}_{1S}$ beyond $40\%$. That $\mathcal{R}_{1S}$ is fixed
already at leading order and barely shifted by the relativistic correction makes it a robust
discriminator, and it selects the smaller mixing angle, in agreement with the leading-order
conclusion of Ref.~\cite{He:2019}. It is also consistent with the mixing angle extracted from our
analyses of the radiative transitions $h_{c}\to\gamma\eta^{(\prime)}$~\cite{Fan:2019sap,He:2020}.

\subsection{Results for the $\psi(2S)$}
\label{subsec:psi2S}

The $\psi(2S)$ amplitudes share most of the structure found for the $J/\psi$. They show a similar
dependence on $m_{q}$ and the DA, the dominant quark amplitude being nearly
insensitive to both, and almost the same hierarchy of the three contributions. What changes markedly from the $J/\psi$ is the size of
the relativistic correction, which is substantially larger. This is intrinsic to a radial excitation,
whose broader, nodal wave function gives a ratio $\nabla^{2}R(0)/R(0)$ about 3 times the
$J/\psi$ value in magnitude, as shown in Eq.~(\ref{eq:lapR}). The $q^{2}$ corrections are correspondingly larger, as
detailed below.

As for the $J/\psi$, we present the $\psi(2S)$ helicity amplitudes in the lattice scheme to show the
size of the relativistic correction. Table~\ref{tab:ampPsi2S} gives the quark, gluon, and QED amplitudes $H^{q}$,
$H^{g}$, and $H^{\rm QED}$ in the $\gamma\eta$ and $\gamma\eta'$ channels, in units of
$10^{-3}~\mathrm{GeV}$. Each amplitude is separated into a leading-order part LO and its $q^{2}$ correction. A single entry is
representative, model~I for the DA and $m_{q}=0.01~\mathrm{GeV}$. The quark
amplitude is again by far the largest. The gluon amplitude is about $3\%$ of it for $\gamma\eta$ and
$9\%$ for $\gamma\etap$, and the QED amplitude is about $20\%$ and $6\%$.

\begin{table}[H]
\caption{Helicity amplitudes of the three contributions to $\psi(2S)\to\gamma\etapp$ in the lattice
mixing scheme, for model~I at $m_{q}=0.01~\mathrm{GeV}$, in units of $10^{-3}~\mathrm{GeV}$. For each
contribution LO is the leading-order helicity amplitude and $q^{2}$ its $q^{2}$ relativistic
correction, the full amplitude being their sum.}
\label{tab:ampPsi2S}
\centering
\begin{ruledtabular}
\begin{tabular}{l c c c c}
 & \multicolumn{2}{c}{$\psi(2S)\to\gamma\eta$} & \multicolumn{2}{c}{$\psi(2S)\to\gamma\etap$} \\
\cline{2-3}\cline{4-5}
Contribution & LO & $q^{2}$ & LO & $q^{2}$ \\ \hline
Quark $H^{q}$     & $-0.55-0.13i$ & $-0.53$ & $-3.04-1.59i$ & $-2.53-1.52i$ \\
Gluon $H^{g}$     & $-0.01$ & $-0.02$ & $-0.19$ & $-0.40$ \\
QED $H^{\rm QED}$ & $-0.22$ & $0$     & $-0.35$ & $0$ \\
\end{tabular}
\end{ruledtabular}
\end{table}

The relativistic correction follows the same pattern as for the $J/\psi$ but is considerably larger,
as Table~\ref{tab:ampPsi2S} shows. The quark term again dominates, and its $q^{2}$ piece is now
nearly as large as the leading order itself, so that $|H^{q}|$ almost doubles in both channels,
compared with the growth of about $40\%$ found for the $J/\psi$. The gluon term, the smallest of the
three, is enhanced most strongly in relative terms, its $q^{2}$ piece reaching about twice the leading
order. The QED term again receives no correction, its single-point $c\bar c$ annihilation leaving no
charmonium relative momentum in the kernel. Summed over the three, the correction enlarges the total
$|H|$ by some $70\%$--$85\%$, roughly twice the effect in the $J/\psi$, in step with the larger ratio
$\nabla^{2}R_{\psi(2S)}(0)/R_{\psi(2S)}(0)$ of Eq.~(\ref{eq:lapR}). Notably, the $q^{2}$ quark
amplitude in the $\gamma\eta$ channel appears in Table~\ref{tab:ampPsi2S} with no imaginary part,
although the leading order carries one. The imaginary part of $H^{q}$ originates
from the internal light-quark and gluon lines going on shell and diminishes as $x=m_{P}^{2}/M^{2}$
decreases, and the absorptive part of the $q^{2}$ correction vanishes faster still, as it is built
from higher powers of these propagators. Both $\psi(2S)$ channels sit at small $x$, and the $\gamma\eta$ one has the
smallest $x$ of all the channels considered in this work, so its absorptive parts are the most
strongly suppressed. Numerically, that of the $q^{2}$ term is at the $10^{-6}~\mathrm{GeV}$ level
and is dropped from the table.

\begin{table}[H]
\caption{Branching ratios $\mathcal{B}(\psi(2S)\rightarrow\gamma\etapp)$ in units of $10^{-4}$, for
model~I at $m_{q}=0.01~\mathrm{GeV}$, in the lattice and $\etap$TFF mixing schemes. Here
$\mathcal{B}_{\rm LO}$ is the leading-order branching ratio, $\mathcal{B}_{q^{2}}$ the one with the
$q^{2}$ relativistic correction included, and $\mathcal{B}_{\rm exp}$ the measured
value~\cite{BESIII:2017geta,ParticleDataGroup:2024}, while the last row gives the ratio
$\mathcal{R}_{2S}$ of the $\gamma\etap$ to the $\gamma\eta$ branching ratio.}
\label{tab:widthsPsi2S}
\centering
\begin{ruledtabular}
\begin{tabular}{l cc cc c}
 & \multicolumn{2}{c}{Lattice} & \multicolumn{2}{c}{$\etap$TFF} & \\
\cline{2-3}\cline{4-5}
Process & $\mathcal{B}_{\rm LO}$ & $\mathcal{B}_{q^{2}}$ & $\mathcal{B}_{\rm LO}$ & $\mathcal{B}_{q^{2}}$ & $\mathcal{B}_{\rm exp}$~\cite{ParticleDataGroup:2024} \\ \hline
$\psi(2S)\rightarrow\gamma\eta$  & $0.07(3)$ & $0.18(7)$ & $0.30(3)$ & $0.76(8)$ & $0.0092(18)$ \\
$\psi(2S)\rightarrow\gamma\etap$ & $1.76(9)$ & $4.71(23)$ & $1.24(7)$ & $3.33(18)$ & $1.24(4)$ \\
$\mathcal{R}_{2S}$ & $23(9)$ & $26(11)$ & $4.2(5)$ & $4.4(6)$ & $135(27)$ \\
\end{tabular}
\end{ruledtabular}
\end{table}

The branching ratios and their ratio $\mathcal{R}_{2S}=\mathcal{B}(\psi(2S)\to\gamma\etap)/\mathcal{B}(\psi(2S)\to\gamma\eta)$ in Table~\ref{tab:widthsPsi2S} show a picture qualitatively unlike that for the $J/\psi$.
The theory overshoots the data in both channels already at leading order, and the relativistic
correction, enlarging each branching ratio by a further factor of about $2.5$, widens the gap. In the
lattice scheme the $\gamma\etap$ rate is comparable to experiment at leading order but reaches almost
4 times the measured value once the correction is included, while the $\gamma\eta$ rate exceeds its
measurement by about a factor of 8 at leading order and by some 20 after the correction. In
the $\etap$TFF scheme the $\gamma\etap$ rate agrees with experiment at leading order and overshoots it
by nearly a factor of 3 with the correction, while the $\gamma\eta$ rate lies more than a factor
of 30 above the data at leading order and close to 2 orders of magnitude above with the
correction. The ratio makes the tension sharpest. The theoretical $\mathcal{R}_{2S}$, $26(11)$ in the
lattice scheme and $4.4(6)$ in the $\etap$TFF one, falls far below the measured $135(27)$, a
discrepancy neither representative mixing scheme can accommodate. The largeness of the measured ratio
stems mainly from the anomalously small $\gamma\eta$ rate.

The overshoot in both channels reflects the
difficulty of the relativistic expansion for the $\psi(2S)$. The $q^{2}$ correction is as large as the
leading order, so the low-order relative-momentum expansion converges poorly, and the node of the $2S$
radial wave function, whose short-distance structure a low-order Taylor expansion cannot capture,
compounds the problem. The same poor convergence was found in our recent study of the three-gluon
decay of the $\psi(2S)$~\cite{He:2026ggg2S}, where the relative correction reaches $2.70$ and the
$q^{2}$-truncated width turns negative, and it is the general behavior of the radial excitation. The
corrections to $\psi(nS)\to\rho\pi$ and to $\psip\to p\bar p$ are of order unity, the corresponding
analyses assigning $\vsq\simeq0.64$ to the $\psip$ against $\simeq0.3$ for the
$\Jpsi$~\cite{Kivel:2022pp,Kivel:2023rp}, and the $v^{2}$ corrections to $\psip$ gluon fragmentation
raise the cross section by factors of $3.3$--$6.6$~\cite{Bertone:2025jex}. Measured against these,
the enhancement found here, which tracks the ratio of the two $\vsq$, is what the velocity expansion
of a radially excited charmonium leads one to expect.
Beyond the expansion itself, the overshoot of the individual branching ratios might in part be laid
to the inputs. The uncertainties in $R_{\psi(2S)}(0)$ and in $\alpha_{s}(\mu)$ are of much the
same size here as for the $\Jpsi$ in Sec.~\ref{subsec:Jpsi}, but they act on the two channels alike,
so that they largely cancel in $\mathcal{R}_{2S}$. What survives is far too small to bring
$\mathcal{R}_{2S}$ near the measured value. The charm mass, fixed in the hard kernel by the
weak-binding approximation of Sec.~\ref{sec:framework}, is a further input to be tested and one to
which the two charmonia respond very differently. Lowering it to the pole-type value
$1.4~\mathrm{GeV}$ used for the charm quark in NRQCD analyses, at the lower end of the constituent
range quoted in Sec.~\ref{subsec:inputs}, and keeping the Cornell wave-function inputs fixed, leaves
the $\Jpsi$ leading-order widths within $7\%$ and raises its $q^{2}$-corrected branching ratios by
$40\%$ in $\gamma\eta$ and $27\%$ in $\gamma\etap$. For the $\psip$ the same change raises the branching
ratios by factors between $2.7$ and $4.3$, the wider spread of internal momentum that enhances the
$q^{2}$ term also sharpening the response of the charm propagator that lies close to its mass shell.
The ratios are far less affected, $\mathcal{R}_{1S}$ moving from $26$ to $24$ and $\mathcal{R}_{2S}$
from $26$ to $18$ in the lattice scheme, and from $4.3$ to $3.8$ and from $4.4$ to $2.8$ in the
$\etap$TFF one, neither of the $\psip$ values approaching the measured ratio. The severe deficit of
$\mathcal{R}_{2S}$ can therefore hardly be explained by the hard mechanism alone, which keeps the two
channels within an order of magnitude of each other and cannot produce so strong a suppression of the
$\gamma\eta$ rate. This suggests that a mechanism beyond the hard perturbative process may also play a
significant role in these decays~\cite{Gerard:2013gya}. In the following we explore the $\etac$-mixing
contribution as one such candidate, although a complete understanding of why the measured
$\mathcal{R}_{2S}$ lies so far above the hard-mechanism prediction remains an open question.

\subsection{Reassessment of the $\etac$-mixing contribution}
\label{subsec:softreassess}

The mechanism analyzed above proceeds by annihilating the $c\bar c$ pair at short
distance into the photon and the partons that build up the light meson. The same decay also receives
a contribution from a second, physically distinct mechanism, in which the $c\bar c$ pair is not
annihilated but survives and enters the final state through the $\etac$ component of the meson. In the picture introduced by Chao~\cite{Chao:1989pi,Chao:1990im},
the physical $\eta$ and $\etap$ each carry a small admixture of the charmonium $\etac$, generated by
the $U_{A}(1)$ anomaly, which couples both the light-quark content and the $\etac$ to two gluons and
lets the two oscillate into each other. The decay $\psi_{n}\rightarrow\gamma\eta^{(\prime)}$ can then
proceed as the magnetic-dipole transition $\psi_{n}\rightarrow\gamma\etac$ on this component. While
the annihilation contribution is fixed by the charmonium wave function at the origin, this transition is
set by the radial overlap of the $\psi_{n}$ and $\etac$ wave functions, which depends on the photon
energy. In the
$\gamma\eta^{(\prime)}$ channels that energy is large and the $\etac$ is probed far off its mass
shell. This large energy confines the overlap to short distances, a small part of the
extended wave functions, so the transition is suppressed, an effect encoded in a radial overlap form
factor. The off-shell $\etac$ then converts into the $\eta^{(\prime)}$ through the two soft gluons of
the anomaly that realize the mixing. In one mechanism the $c\bar c$ pair
annihilates and in the other it survives. Yet both reach
the same $\gamma\eta^{(\prime)}$ final state through different components of the $\eta^{(\prime)}$, so
their amplitudes add coherently.

To gauge the size of the $\etac$-mixing contribution we now construct its amplitude, whose structure is fixed
completely by the quantum numbers. Both
$\psi_{n}\rightarrow\gamma\eta^{(\prime)}$ and $\psi_{n}\rightarrow\gamma\etac$ are
$1^{--}\rightarrow\gamma\,0^{-+}$ transitions and carry the single Lorentz structure
$\epsilon_{\mu\nu\lambda\sigma}K^{\lambda}k^{\sigma}$, so the $\etac$-mixing amplitude reads
\begin{equation}
\mathcal{M}^{\etac}(\psi_{n}\rightarrow\gamma\eta^{(\prime)})
=c^{\eta^{(\prime)}}_{\etac}\,g_{\psi_{n}\gamma\etac}\,F_{n}(k_{\eta^{(\prime)}})\,
\epsilon_{\mu\nu\lambda\sigma}K^{\lambda}k^{\sigma}\,
\eps^{\mu}(K)\,\eps^{*\nu}(k),
\label{eq:Metacmix}
\end{equation}
where the form factor is evaluated at the photon energy
$k_{\eta^{(\prime)}}=(M^{2}-m_{\eta^{(\prime)}}^{2})/(2M)$ of the channel in the $\psi_{n}$ rest
frame. The amplitude is built from three quantities, the $\etac$
admixture $c^{\eta^{(\prime)}}_{\etac}$, the radial overlap form factor $F_{n}$, and the effective
coupling $g_{\psi_{n}\gamma\etac}$, which we now fix in turn.

The admixture is taken from the charm-content analysis of Ref.~\cite{Ali:1997ex}. There the $c\bar c$
pair created by the axial-vector current converts into two gluons through the charm-quark loop. For
the heavy charm this loop is a short-distance object and is integrated out, and the matrix element of
the resulting effective interaction in the $\eta^{(\prime)}$ defines the charm decay constant of the
light pseudoscalar $f^{c}_{\eta^{(\prime)}}$.\footnote{Explicitly
$f^{c}_{\eta^{(\prime)}}=-\Delta i_{5}(m_{\eta^{(\prime)}}^{2}/m_{c}^{2})\,f^{u}_{\eta^{(\prime)}}$,
with $\Delta i_{5}(z)=-1+\big[\pi-2\arctan\sqrt{4/z-1}\,\big]^{2}/z\approx z/12$ at small
$z$~\cite{Ali:1997ex}, so the charm content is suppressed by $m_{\eta^{(\prime)}}^{2}/m_{c}^{2}$. The
charm mass here is taken to be the current mass,
$m_{c}(m_{c})=1.273(5)~\mathrm{GeV}$~\cite{ParticleDataGroup:2024}.}
The admixture follows as $c^{\eta^{(\prime)}}_{\etac}=f^{c}_{\eta^{(\prime)}}/f_{\etac}$, where the
$\etac$ decay constant $f_{\etac}=0.387(7)~\mathrm{GeV}$ is taken from the lattice determination of
Ref.~\cite{Becirevic:2013bsa}. In the lattice mixing scheme this
gives $c^{\eta}_{\etac}=-0.00310(15)$ and $c^{\etap}_{\etac}=-0.00820(47)$, in line with the values
of Ref.~\cite{Beneke:2002jn} and about half those extracted phenomenologically in the
FKS analysis~\cite{Feldmann:1998vh}.

The form factor is the radial overlap of the initial charmonium with the $\etac$ at photon energy
$k$~\cite{Brambilla:2005zw},
\begin{equation}
F_{n}(k)=\int_{0}^{\infty}dr\,r^{2}\,R_{10}(r)\,R_{n0}(r)\,j_{0}(kr/2),
\label{eq:FFdef}
\end{equation}
with $R_{10}$ and $R_{n0}$ the radial wave functions of the $\etac$ and the initial charmonium,
$n=1$ the allowed $1S\rightarrow1S$ transition of the $J/\psi$, and $n=2$ the hindered
$2S\rightarrow1S$ one of the $\psi(2S)$, for which $F_{2}(0)=0$ by orthogonality. For these radial
wave functions we adopt a simple but analytic approximation, the harmonic-oscillator form, which
gives
\begin{equation}
F_{1}(k)=e^{-k^{2}a^{2}/16},\qquad
F_{2}(k)=\frac{\sqrt6}{48}\,(ka)^{2}\,e^{-k^{2}a^{2}/16}.
\label{eq:FFHO}
\end{equation}
The oscillator length $a$, the essential nonperturbative input, is related to the charmonium size by
the harmonic-oscillator identity $\langle r^{2}\rangle_{1S}=3a^{2}/2$. We fix it from the $1S$
root-mean-square radius of the same Cornell potential~\cite{Eichten:2019gig,Eichten:1978tg} that
determines the wave functions at the origin in Sec.~\ref{subsec:inputs}. Solving the radial
Schr\"odinger equation with this potential gives
$\langle r^{2}\rangle^{1/2}_{1S}=0.39~\mathrm{fm}$, and hence $a=1.62~\mathrm{GeV}^{-1}$. This radius
agrees with recent potential-model determinations of the charmonium size~\cite{Akbar:2023rms}.
Against the numerical form factors of the Cornell potential the oscillator forms deviate by no more
than a few percent, and at the large photon momenta of the $\gamma\eta^{(\prime)}$ channels the two
are nearly identical. Their energy dependence, $k^{3}|F_{1}|^{2}$ for the allowed and $k^{7}$ for the
hindered transition, is moreover consistent with the photon-energy dependence of the $\etac$ line
shapes observed by CLEO-c and BESIII in
$\psi_{n}\rightarrow\gamma\etac$ away from the resonance peak~\cite{CLEO:2008etac,BESIII:2011etac}.

The effective coupling is anchored to the measured rates. The same vertex with
$c^{\eta^{(\prime)}}_{\etac}\rightarrow1$ describes the on-shell transition
$\psi_{n}\rightarrow\gamma\etac$, whose width
$\Gamma=|g_{\psi_{n}\gamma\etac}|^{2}|F_{n}(k_{\etac})|^{2}k_{\etac}^{3}/(12\pi)$, with $k_{\etac}$
the photon energy of this on-shell transition, inverts to
\begin{equation}
|g_{\psi_{n}\gamma\etac}|=\frac{1}{|F_{n}(k_{\etac})|}
\sqrt{\frac{12\pi\,\Gamma^{\rm exp}(\psi_{n}\rightarrow\gamma\etac)}{k_{\etac}^{3}}},
\label{eq:gcoupling}
\end{equation}
where $\Gamma^{\rm exp}(\psi_{n}\rightarrow\gamma\etac)$ is the measured
width~\cite{ParticleDataGroup:2024}. Determining the coupling from the measured width is necessary
because the magnetic-dipole transitions are not yet under full theoretical control. For
$J/\psi\rightarrow\gamma\etac$ the nonrelativistic prediction overshoots the measured rate by a
factor of 2--3, and the relativistic corrections of relative order $v^{2}$ restore agreement
with experiment within sizable uncertainties~\cite{Brambilla:2005zw}. For the hindered
$2S\rightarrow1S$ transition, dominated by relativistic corrections and complicated further by the
proximity of the $\psi(2S)$ to the open-charm threshold, a reliable theoretical prediction is more
difficult still~\cite{Brambilla:2005zw,Radford:2008,Deng:2016stx,Delaney:2023}. Anchoring to data
absorbs this difficulty but fixes only the magnitude of the coupling, leaving its phase
undetermined.

Applying the helicity projector to Eq.~(\ref{eq:Metacmix}) gives the
$\etac$-mixing helicity amplitude $H^{\etac}$. Let $H^{\rm ann}\equiv H^{q}+H^{g}+H^{\rm QED}$ denote
the summed helicity amplitude of the quark, gluon, and QED contributions analyzed before. As argued
at the beginning of this section, $H^{\etac}$ adds coherently to $H^{\rm ann}$, giving the total
amplitude
\begin{equation}
H=H^{\rm ann}+H^{\etac}.
\label{eq:cohsum}
\end{equation}
Here $H^{\etac}=|H^{\etac}|\,e^{i\delta_{nS}}$, with $\delta_{1S}$ and $\delta_{2S}$ treated as two free
parameters, each common to the $\gamma\eta$ and $\gamma\etap$ channels of its charmonium, and fixed
by a best fit to experiment.

The fit is anchored to the two measured $\gamma\eta$ branching ratios, which we carry out in the
lattice scheme as a representative case, and returns $\delta_{1S}\simeq192^{\circ}$ and
$\delta_{2S}\simeq18^{\circ}$. Table~\ref{tab:cohfit} collects the four resulting branching ratios. The interference
brings the $\psi(2S)$ into agreement with experiment in both channels and substantially improves the
$J/\psi$, whose $\gamma\eta$ and $\gamma\etap$ rates come to about half and twice the measured
values. It is worth noting that the two fitted phases differ by nearly $180^{\circ}$, which
means the $\etac$-mixing amplitudes of the two charmonia are almost opposite in sign. This relative
sign finds a natural interpretation in the magnetic-dipole dynamics, where the relativistic
correction is opposite in sign to the leading order~\cite{Brambilla:2005zw,Radford:2008,Deng:2016stx}.
The leading order dominates the $J/\psi$ amplitude but vanishes for the $\psi(2S)$ by the $1S$--$2S$
orthogonality, leaving the $\psi(2S)$ amplitude as the correction alone, opposite in sign to that of
the $J/\psi$.

\begin{table}[H]
\caption{Branching ratios of $\psi_{n}\rightarrow\gamma\etapp$ from the coherent
amplitude at the best-fit phases $\delta_{1S}=192^{\circ}$ and
$\delta_{2S}=18^{\circ}$, in the lattice mixing scheme, compared with the measured
values~\cite{ParticleDataGroup:2024}. The
$J/\psi$ entries are in units of $10^{-3}$ and the $\psi(2S)$ entries in $10^{-4}$.}
\label{tab:cohfit}
\centering
\begin{ruledtabular}
\begin{tabular}{l c c}
Process & $\mathcal{B}_{\rm fit}$ & $\mathcal{B}_{\rm exp}$ \\ \hline
$J/\psi\rightarrow\gamma\eta$    & $0.65(14)$   & $1.090(13)$ \\
$J/\psi\rightarrow\gamma\etap$   & $9.34(52)$   & $5.28(6)$ \\
$\psi(2S)\rightarrow\gamma\eta$  & $0.0092(47)$ & $0.0092(18)$ \\
$\psi(2S)\rightarrow\gamma\etap$ & $1.97(29)$   & $1.24(4)$ \\
\end{tabular}
\end{ruledtabular}
\end{table}

Encouraging as this is, the coherent results obtained above should be regarded as tentative. We have deliberately
fixed the essential nonperturbative inputs of both contributions, the wave function at the origin of
$H^{\rm ann}$ and the radial overlap of the $\etac$-mixing one, from the same Cornell
potential, so as to minimize the model dependence of their coherence. A residual dependence
nonetheless remains. Both contributions depend on the mixing parameters, and in the $\gamma\eta$
channel $H^{\rm ann}$ is especially sensitive to them through the near cancellation in
$\sqrt2 f_{q}\cos\phi-f_{s}\sin\phi$. The relativistic expansion, moreover, converges poorly for the
$\psi(2S)$, which degrades the reliability of $H^{\rm ann}$ there. These limitations make a
precise extraction of the interference difficult, and a more reliable determination of the relative
phases is left to a dedicated future study.

\section{Summary}
\label{sec:summary}

We have computed the OZI-forbidden radiative decays $J/\psi,\psi(2S)\rightarrow\gamma\eta^{(\prime)}$
in perturbative QCD, resolving the amplitude into its quark-antiquark, two-gluon, and QED
contributions and including the leading $q^{2}$ relativistic correction to each. A central result is
the remarkable stability of these amplitudes against the soft inputs of the hard mechanism. The
dominant quark-antiquark contribution is almost completely insensitive to the DA and to the light-quark mass, varying by less than $2\%$ across three
DA models and by less than $1\%$ over the light-quark-mass range, while the
smaller two-gluon and QED contributions carry only a mild DA dependence and are
likewise nearly independent of the light-quark mass. This robustness holds at leading order and
survives the $q^{2}$ correction, confirming the leading-order finding of our earlier
work~\cite{He:2019,He:2020jvj} and extending it to relative order $q^{2}$. Recent lattice studies have raised as an open question
the light-quark-mass dependence of $J/\psi\rightarrow\gamma\eta^{(\prime)}$~\cite{HadronSpectrum:2025long,HadronSpectrum:2025letter},
while our perturbative analysis, borne out numerically, points to a weak dependence. This
insensitivity to the soft inputs makes our predictions for the process correspondingly reliable.

The $q^{2}$ correction enhances the amplitude in every channel, but by amounts that differ markedly
between the two charmonia. For the $J/\psi$ this enhancement is controlled, about $40\%$ at the
amplitude level and in line with the charmonium mean squared velocity $\langle v^{2}\rangle\simeq0.3$,
and it roughly doubles the branching ratios and so narrows their sizable shortfall from the measured
values. For the $\psi(2S)$ the enhancement is about twice as large as for the $J/\psi$. There the
$q^{2}$ term is comparable to the leading order, a poor convergence that undermines the
relative-momentum expansion and makes a quantitative account of the $\psi(2S)$ relativistic
correction difficult.

Comparing with experiment in two representative mixing schemes, the lattice and the $\etap$TFF, we
find the relativistic correction essential in approaching the measured $J/\psi$ branching ratios,
while their ratio $\mathcal{R}_{1S}$ favors the smaller mixing angle. Both this ratio and the
$\gamma\eta$ branching ratio are highly sensitive to the mixing angle, so they discriminate sharply
between the schemes. For the $\psi(2S)$ the annihilation contribution overshoots both channels, the
$\gamma\etap$ rate moderately and the anomalously small $\gamma\eta$ rate by far more. As a candidate
resolution we examined the interference with the $\etac$-mixing contribution, whose helicity
amplitude, normalized to the measured $\psi_{n}\rightarrow\gamma\etac$ rates and suppressed by a
radial overlap form factor, is comparable in size to the annihilation one in the $\gamma\eta$
channel. Added coherently, it can bring both $\psi(2S)$ channels back to their measured values. The
strong mixing-parameter sensitivity of the $\gamma\eta$ rates and of the ratios, however, ties this
interference closely to the choice of mixing scheme and makes a clean extraction of the coherence
between the two contributions difficult. Sharpening this extraction will require better-constrained
mixing parameters and a firmer handle on the phase of the $\etac$-mixing amplitude. The poor
convergence of the $\psi(2S)$ relativistic expansion, in turn, calls for retaining the full
relative-momentum dependence of the hard kernel rather than its $q^{2}$ truncation. We leave both to
future work.

\begin{acknowledgments}
The work of J.-K.~H. is supported by the National Natural Science Foundation of China under
Grant No.~12305086 and the Project of Science and Technology Research Program of the Hubei
Provincial Department of Education under Grant No.~Q20222504. The work of C.-J.~F. is supported
by the Hubei Provincial Natural Science Foundation Youth Project under Grant No.~2024AFB151.
\end{acknowledgments}

\section*{Data availability}
There are no publicly available research data or software supporting this manuscript.
Requests for further information or data should be sent to the authors.

\appendix

\section{Explicit forms of the higher denominator-power numerators $N^{X}_{a,b}$ in Eq.~(\ref{eq:integq2})}
\label{app:NXab}

The higher denominator-power numerators $N^{X}_{a,b}$ ($a+b>2$) of the $q^{2}$-correction kernel
$I^{q}_{2}(u)$ are listed below in the weak-binding approximation $m_{c}=M/2$, with $p\equiv K-k$
the $\eta^{(\prime)}$ momentum,
\begin{align}
N^{A}_{1,2}={}& \frac{32 i}{3 M}\Big[-\,\frac{2 M^2\,(k\!\cdot\!k_{1})^{2}\,(k_{1}\!\cdot\!K)}{k\!\cdot\!K} + \frac{2 M^2\,(k\!\cdot\!k_{1})^{3}}{k\!\cdot\!K} - 2\,(k\!\cdot\!k_{1})\,(k_{1}\!\cdot\!K)\,(k\!\cdot\!K) \nonumber \\
    &\quad  - 2\,k_{1}^{2}\,(k_{1}\!\cdot\!K)\,(k\!\cdot\!K) - 2\,k_{1}^{2}\,(k\!\cdot\!k_{1})\,(k\!\cdot\!K) - M^2\,(k\!\cdot\!k_{1})\,(k_{1}\!\cdot\!K) \nonumber \\
    &\quad  + 4 M^2\,(k\!\cdot\!k_{1})^{2} + 4 M^2\,k_{1}^{2}\,k\!\cdot\!k_{1} -\frac{M^4\,(k\!\cdot\!k_{1})^{2}}{k\!\cdot\!K} \nonumber \\
    &\quad  + k_{1}^{2}\,(k\!\cdot\!K)^{2} + 2 M^2\,k_{1}^{2}\,k\!\cdot\!K\,\Big].
\label{eq:NXA12}
\end{align}

\begin{align}
N^{A}_{1,3}={}& \frac{128 i}{3 M}\Big[-\,(k\!\cdot\!k_{1})\,(k_{1}\!\cdot\!K)^{3} + (k\!\cdot\!k_{1})^{2}\,(k_{1}\!\cdot\!K)^{2} + k_{1}^{2}\,(k_{1}\!\cdot\!K)^{2}\,(k\!\cdot\!K) \nonumber \\
    &\quad  + M^2\,k_{1}^{2}\,(k\!\cdot\!k_{1})\,(k_{1}\!\cdot\!K) - M^2\,k_{1}^{2}\,(k\!\cdot\!k_{1})^{2} - M^2\,(k_{1}^{2})^{2}\,k\!\cdot\!K\,\Big].
\label{eq:NXA13}
\end{align}

\begin{align}
N^{A}_{2,1}={}& \frac{32 i}{3 M}\Big[-\,4\,(k\!\cdot\!k_{1})\,(k_{1}\!\cdot\!K)\,(k\!\cdot\!K) + 4\,(k\!\cdot\!k_{1})^{2}\,(k\!\cdot\!K) - 3 M^2\,(k\!\cdot\!k_{1})\,(k_{1}\!\cdot\!K) \nonumber \\
    &\quad  + 3 M^2\,(k\!\cdot\!k_{1})^{2} + 4\,k_{1}^{2}\,(k\!\cdot\!K)^{2} + 3 M^2\,k_{1}^{2}\,k\!\cdot\!K\,\Big].
\label{eq:NXA21}
\end{align}

\begin{align}
N^{A}_{2,2}={}& \frac{128 i}{3 M}\Big[(k\!\cdot\!k_{1})\,(k_{1}\!\cdot\!K)^{2}\,(k\!\cdot\!K) - (k\!\cdot\!k_{1})^{2}\,(k_{1}\!\cdot\!K)\,(k\!\cdot\!K) - M^2\,(k\!\cdot\!k_{1})^{2}\,(k_{1}\!\cdot\!K) \nonumber \\
    &\quad  + M^2\,(k\!\cdot\!k_{1})^{3} - k_{1}^{2}\,(k_{1}\!\cdot\!K)\,(k\!\cdot\!K)^{2} + M^2\,k_{1}^{2}\,(k\!\cdot\!k_{1})\,(k\!\cdot\!K)\,\Big].
\label{eq:NXA22}
\end{align}

\begin{align}
N^{A}_{3,1}={}& \frac{128 i}{3 M}\Big[-\,(k\!\cdot\!k_{1})\,(k_{1}\!\cdot\!K)\,(k\!\cdot\!K)^{2} + (k\!\cdot\!k_{1})^{2}\,(k\!\cdot\!K)^{2} + k_{1}^{2}\,(k\!\cdot\!K)^{3}\,\Big].
\label{eq:NXA31}
\end{align}

\begin{align}
N^{C}_{1,2}={}& \frac{32 i}{3 M}\Big[8\,(k\!\cdot\!k_{1})\,(k_{1}\!\cdot\!K)^{2} - 8\,(k\!\cdot\!k_{1})^{2}\,(k_{1}\!\cdot\!K) - 4\,k_{1}^{2}\,(k\!\cdot\!k_{1})\,(k_{1}\!\cdot\!K) \nonumber \\
    &\quad  - \frac{4 M^2\,(k\!\cdot\!k_{1})^{2}\,(k_{1}\!\cdot\!K)}{k\!\cdot\!K} + \frac{4 M^2\,(k\!\cdot\!k_{1})^{3}}{k\!\cdot\!K} + \frac{2 M^2\,k_{1}^{2}\,(k\!\cdot\!k_{1})^{2}}{k\!\cdot\!K} \nonumber \\
    &\quad  + 6\,(k\!\cdot\!k_{1})\,(k_{1}\!\cdot\!K)\,(k\!\cdot\!K) - 4\,k_{1}^{2}\,(k_{1}\!\cdot\!K)\,(k\!\cdot\!K) + 4\,k_{1}^{2}\,(k\!\cdot\!k_{1})\,(k\!\cdot\!K) \nonumber \\
    &\quad  + 2\,(k_{1}^{2})^{2}\,k\!\cdot\!K + 4 M^2\,(k\!\cdot\!k_{1})\,(k_{1}\!\cdot\!K) - 3 M^2\,(k\!\cdot\!k_{1})^{2} \nonumber \\
    &\quad  - \frac{2 M^4\,(k\!\cdot\!k_{1})^{2}}{k\!\cdot\!K} - 3\,k_{1}^{2}\,(k\!\cdot\!K)^{2} - 2 M^2\,k_{1}^{2}\,k\!\cdot\!K\,\Big].
\label{eq:NXC12}
\end{align}

\begin{align}
N^{C}_{1,3}={}& \frac{128 i}{3 M}\Big[2\,(k\!\cdot\!k_{1})\,(k_{1}\!\cdot\!K)^{3} - \frac{M^2\,(k\!\cdot\!k_{1})^{2}\,(k_{1}\!\cdot\!K)^{2}}{k\!\cdot\!K} + 4\,(k\!\cdot\!k_{1})\,(k_{1}\!\cdot\!K)^{2}\,(k\!\cdot\!K) \nonumber \\
    &\quad  - k_{1}^{2}\,(k_{1}\!\cdot\!K)^{2}\,(k\!\cdot\!K) - 6 M^2\,(k\!\cdot\!k_{1})^{2}\,(k_{1}\!\cdot\!K) - 2 M^2\,k_{1}^{2}\,(k\!\cdot\!k_{1})\,(k_{1}\!\cdot\!K) \nonumber \\
    &\quad  + \frac{2 M^4\,(k\!\cdot\!k_{1})^{3}}{k\!\cdot\!K} + \frac{M^4\,k_{1}^{2}\,(k\!\cdot\!k_{1})^{2}}{k\!\cdot\!K} + 2\,(k\!\cdot\!k_{1})\,(k_{1}\!\cdot\!K)\,(k\!\cdot\!K)^{2} \nonumber \\
    &\quad  - 2\,k_{1}^{2}\,(k_{1}\!\cdot\!K)\,(k\!\cdot\!K)^{2} - M^2\,(k\!\cdot\!k_{1})^{2}\,(k\!\cdot\!K) + 2 M^2\,k_{1}^{2}\,(k\!\cdot\!k_{1})\,(k\!\cdot\!K) \nonumber \\
    &\quad  + M^2\,(k_{1}^{2})^{2}\,k\!\cdot\!K - k_{1}^{2}\,(k\!\cdot\!K)^{3}\,\Big].
\label{eq:NXC13}
\end{align}

\begin{align}
N^{C}_{2,2}={}& \frac{128 i}{3 M}\Big[2\,(k\!\cdot\!k_{1})\,(k_{1}\!\cdot\!K)^{3} - \frac{M^2\,(k\!\cdot\!k_{1})^{2}\,(k_{1}\!\cdot\!K)^{2}}{k\!\cdot\!K} + 2\,(k\!\cdot\!k_{1})\,(k_{1}\!\cdot\!K)^{2}\,(k\!\cdot\!K) \nonumber \\
    &\quad  - k_{1}^{2}\,(k_{1}\!\cdot\!K)^{2}\,(k\!\cdot\!K) - 3 M^2\,(k\!\cdot\!k_{1})^{2}\,(k_{1}\!\cdot\!K) - 2 M^2\,k_{1}^{2}\,(k\!\cdot\!k_{1})\,(k_{1}\!\cdot\!K) \nonumber \\
    &\quad  + \frac{M^4\,(k\!\cdot\!k_{1})^{3}}{k\!\cdot\!K} + \frac{M^4\,k_{1}^{2}\,(k\!\cdot\!k_{1})^{2}}{k\!\cdot\!K} - k_{1}^{2}\,(k_{1}\!\cdot\!K)\,(k\!\cdot\!K)^{2} \nonumber \\
    &\quad  + M^2\,k_{1}^{2}\,(k\!\cdot\!k_{1})\,(k\!\cdot\!K) + M^2\,(k_{1}^{2})^{2}\,k\!\cdot\!K\,\Big].
\label{eq:NXC22}
\end{align}
The remaining numerators follow from the Bose symmetry under the gluon interchange
$k_{1}\rightarrow p-k_{1}$, which gives $N^{B}_{a,b}=N^{A}_{a,b}(k_{1}\rightarrow p-k_{1})$ and
$N^{C}_{a,b}=N^{C}_{b,a}(k_{1}\rightarrow p-k_{1})$.

\bibliography{refs}

\end{document}